\documentclass{vgtc}                          % final (conference style)
\ifpdf%                                % if we use pdflatex
  \pdfoutput=1\relax                   % create PDFs from pdfLaTeX
  \usepackage{graphicx}                % allow us to embed graphics files
  \DeclareGraphicsExtensions{.pdf,.png,.jpg,.jpeg} % for pdflatex we expect .pdf, .png, or .jpg files
\else%                                 % else we use pure latex
  \usepackage{graphicx}                % allow us to embed graphics files
  \DeclareGraphicsExtensions{.eps}     % for pure latex we expect eps files
\fi%

\graphicspath{{figures/}{pictures/}{images/}{./}} % where to search for the images

\usepackage{microtype}                 % use micro-typography (slightly more compact, better to read)
\PassOptionsToPackage{warn}{textcomp}  % to address font issues with \textrightarrow
\usepackage{textcomp}                  % use better special symbols
\usepackage{mathptmx}                  % use matching math font
\usepackage{times}                     % we use Times as the main font
\usepackage{cite}                      % needed to automatically sort the references
\usepackage{tabu}                      % only used for the table example
\usepackage{booktabs}                  % only used for the table example
\usepackage{amsmath}
\usepackage{amsfonts}
\usepackage{adjustbox}
\usepackage{color, colortbl}
\usepackage{multirow}
\onlineid{0}

\vgtccategory{Research}

\vgtcinsertpkg

\newcommand{\hide}[1]{}

\def\RR{\mathbb{R}}

\title{Volume Rendering of Human Hand Anatomy}

\author{
     Jingtao Huang \thanks{e-mail: jingtaoh@usc.edu}\\ %
     \scriptsize USC 
\and Bohan Wang \thanks{e-mail: bohanwan@mit.edu}\\ %
     \scriptsize MIT
\and Zhiyuan Gao \thanks{e-mail: gaozhiyu@usc.edu}\\%
     \scriptsize USC
\and Mianlun Zheng \thanks{e-mail: mianlunz@usc.edu}\\ %
     \scriptsize USC
\and George Matcuk \thanks{e-mail: matcuk@gmail.com}\\%
     \scriptsize Cedars-Sinai Hospital
\and Jernej Barbi\v{c}\thanks{e-mail: jnb@usc.edu}\\ %
     \scriptsize USC
}

\teaser{
  \centering
  \includegraphics[width=1.0\hsize]{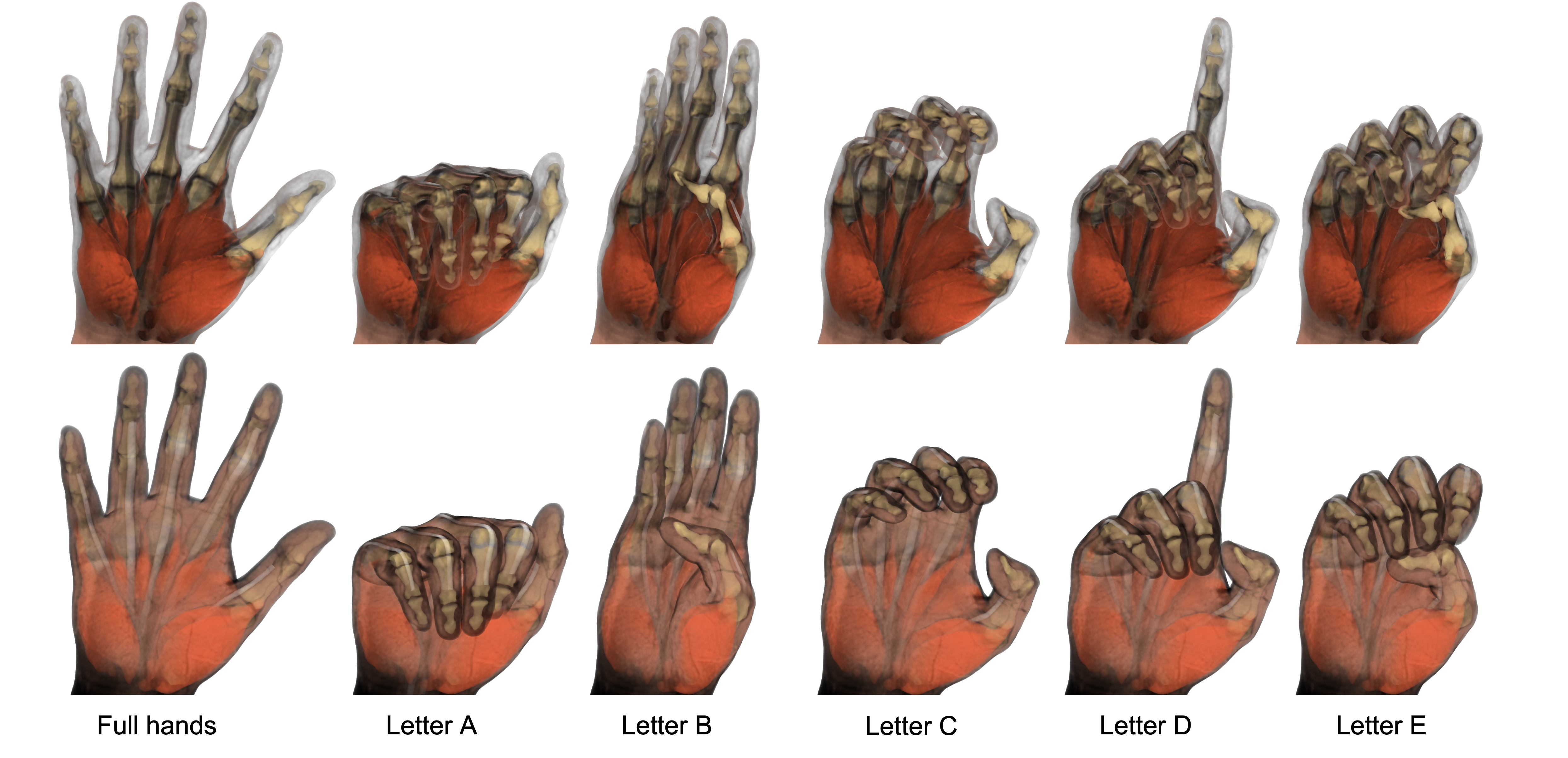}
  \caption{\textbf{"Volume-rendered MRI images of the human hand with our transfer functions"}. The shown hand
poses are from the American Sign Language. The top and bottom rows are rendered using
our interior-emphasized (top) and fat-emphasized style (bottom) transfer functions, respectively.}
  \label{fig:teaser}
}

\abstract{
We study the design of transfer functions for volumetric rendering
of magnetic resonance imaging (MRI) datasets of human hands.
Human hands are anatomically complex, containing various organs within a limited space, 
which presents challenges for volumetric rendering.
We focus on hand musculoskeletal organs
because they are volumetrically the largest inside the hand, 
and most important for the hand's main function, namely manipulation of objects.
While volumetric rendering is a mature field, the choice of the transfer
function for the different organs is arguably just as important as the choice
of the specific volume rendering algorithm; we demonstrate that it significantly influences the 
clarity and interpretability of the resulting images. 
We assume that the hand MRI scans have already been segmented into the different
organs (bones, muscles, tendons, ligaments, subcutaneous fat, etc.).
Our method uses the hand MRI volume data, and the geometry of 
its inner organs and their known segmentation, 
to produce high-quality volume rendering images of the hand, 
and permits fine control over the appearance of each tissue. 
We contribute two families of transfer functions to emphasize different hand tissues of interest, 
while preserving the visual context of the hand. We also discuss and reduce 
artifacts present in standard volume ray-casting of human hands.
We evaluate our volumetric rendering on five challenging hand motion sequences. 
Our experimental results demonstrate that our method improves hand anatomy visualization, 
compared to standard surface and volume rendering techniques.
} % end of abstract

\CCScatlist{
  \CCScatTwelve{Human Hand Anatomy}{Visu\-al\-iza\-tion}{Visu\-al\-iza\-tion techniques}{Volume Rendering};
  \CCScatTwelve{Human Hand Anatomy}{Visu\-al\-iza\-tion}{Visualization design and evaluation methods}{}
}

\definecolor{bone}{RGB}{224, 214, 145}
\definecolor{skin}{RGB}{177, 122, 101}
\definecolor{muscle}{RGB}{255, 98, 56}
\definecolor{tendon}{RGB}{255, 255, 255}
\definecolor{ligament}{RGB}{170, 170, 170}

\begin{document}

%% The ``\maketitle'' command must be the first command after the
%% ``\begin{document}'' command. It prepares and prints the title block.

%% the only exception to this rule is the \firstsection command
\firstsection{Introduction}

\maketitle

%% \section{Introduction} %for journal use above \firstsection{..} instead

The human hand is a vitally important part of the human body. 
The hand is capable of a wide range of precise motions due to its delicate and complex anatomy. 
Each hand consists of 27 bones, 34 muscles, over 100 ligaments and tendons, and numerous blood vessels and nerves, 
all confined into a small volumetric space. Visualizing the hand anatomy can help us understand 
its underlying structure and functionality, including understanding how the hand
volumetrically moves and deforms under realistic hand motions.
Unlike surface (skin) rendering, volumetric rendering of magnetic resonance imaging (MRI)
and computerized tomography (CT) datasets can visualize the underlying interior anatomy. 
MRI scanning has an advantage over CT in that it involves no ionizing radiation, 
and produces better contrast on soft tissues. Because hand consists of many soft 
tissues, we use MRI in our work. 

Volumetric rendering is a well-understood process
in visualization, computer graphics and medical imaging. However, executing a volumetric rendering algorithm to
visualize a hand MRI is only half of the ``story''. Equally important for output image quality is the selection 
of the transfer function that maps MRI signal intensities to colors and opacities.
In our work, we focus on the latter problem, namely design of MRI transfer functions for the human hand. 
The MRI signal is a scalar-valued quantity available at points of 
a regular 3D grid at some resolution (typically around 1mm).
As such, the MRI signal alone has no color or opacity; it is a simple
grayscale value, and there is no obvious a priori way of mapping it to colors. 
By opacity, we mean the standard transparency measure in computer graphics,
i.e., zero opacity implies a fully transparent object, whereas opacity=1
is a fully opaque object; and the in-between values indicate partial
transparency. A transfer function maps the MRI intensity value
to optical properties, namely the (R, G, B) color and opacity.
The transfer function is a critical component of volume rendering.
Its selection is highly non-trivial for biological tissues,
and significantly affects the clarity and interpretability of the output image.

In our work, we investigate how to select spatially-varying per-tissue transfer
functions to produce high-quality volume renderings of the human hand.
We design transfer functions that are suitable for hand musculosketal tissues, 
including bones, muscles, tendons, joint ligaments and fat. 
We do not investigate segmentation and assume that the hand anatomy 
has already been segmented into the different organs, using existing methods.
We apply our novel transfer functions to a standard volume rendering method, namely 
volume ray casting~\cite{Kruger:2003:RAY-CASTING} (referred to as ``ray casting'' in our work). 
We extend ray casting to be aware of both the MRI volume data,
as well as the geometry of the internal organs, producing high-quality volume renders 
that smoothly and clearly display the interior anatomy.
We also address artifacts present in standard volume ray-casting of human hands, namely 
staircase/voxelization artifacts and wood-grain artifacts.
We test our volumetric rendering on a hand animation dataset consisting of
a temporal sequence of MRI images with matching organ geometric shapes. We 
compare our results with renders produced by both standard surface-based 
rendering techniques and volume rendering methods of prior work.

\section{Related Work}
\label{sec:related work}

\subsection{Volume Rendering}

In general, volume rendering methods can be coarsely classified into two categories: indirect volume rendering (IVR), and direct volume rendering (DVR). IVR or ``surface rendering'' fits geometric primitives from the volume data and then renders these primitives. Methods in this category include iso-surfacing \cite{Lorensen:1987:MC} and frequency domain rendering \cite{Totsuka:1993:FDVR}.
DVR or simply ``volume rendering'' renders the volume data by directly forming ``primitives'' from it,
properly superimposing them, optionally combined with segmentation.
There are many volume rendering algorithms. Splatting~\cite{Westover:1991:Splatting} works by virtually ``throwing'' the voxels onto the image plane, with each voxel projecting a ``splat'' in the image plane. Shear-warp rendering~\cite{Cameron:1992:Shear-warp,Lacroute:1994:Shear-warp} uses a projection to form a distorted intermediate image, and a 2D warp to produce the final image. Texture slicing~\cite{Cabral:1994:VRTMH,Van:1996:DVRS3DT} employs GPU hardware to render volume data represented by a 2D or 3D texture. In volume ray casting~\cite{Levoy:1988:DSVD}\cite{Kruger:2003:RAY-CASTING} (Section~\ref{sec:volumeRayCasting}), the camera rays are cast into the MRI volume. Optical properties at sample positions are then integrated along these rays, and the results are the pixels that contribute to the final image. 
Of all the volume rendering algorithms, volume ray casting has seen the largest amount of publications 
and achieves images of highest quality~\cite{Meissner:2000:PEVRA}. 

\subsection{Design of Transfer Functions} 
\label{sec:transfer-functions}
  
\begin{figure}
    \centering
    \includegraphics[width=1.0\hsize]{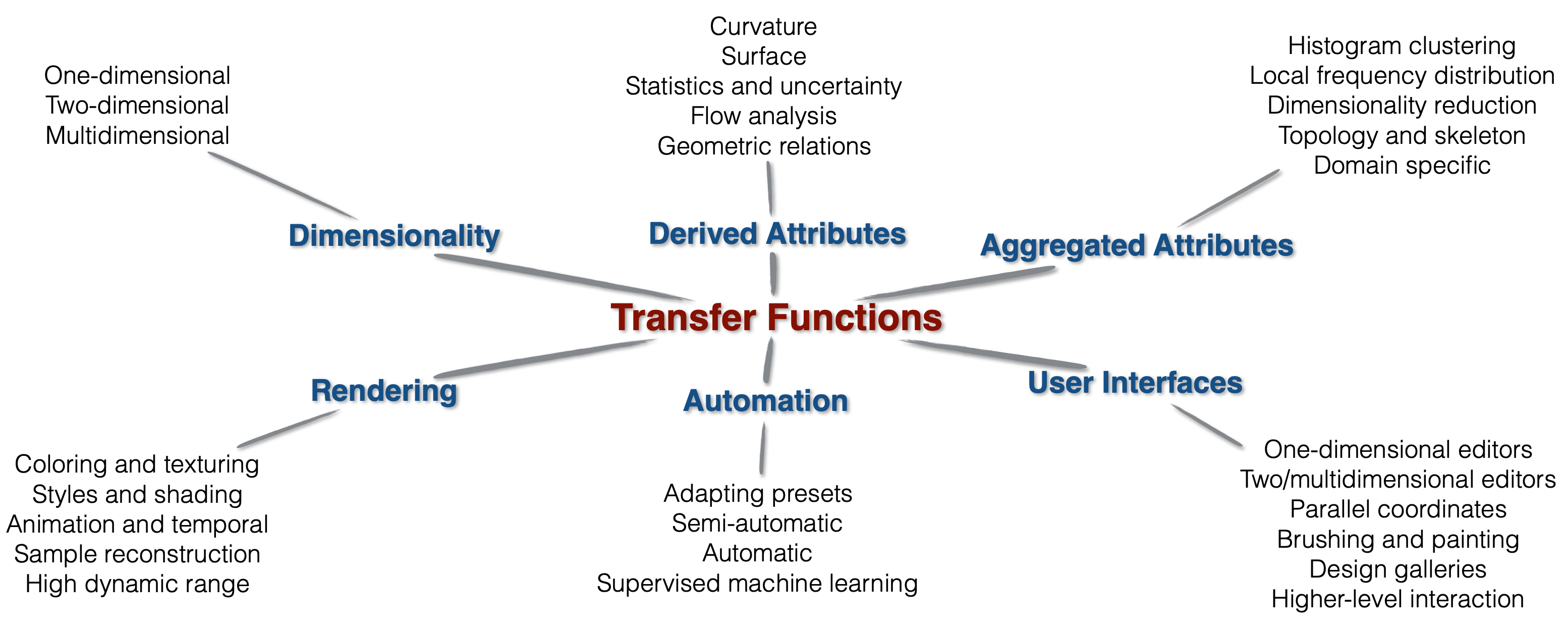}
    \caption[The classification of transfer functions.]{The classification of transfer functions. © 2022 John Wiley and Sons. Reproduced, with permission, from Ljung et al.~\cite{Ljung:2016:START-TF}.}
    \label{fig:TF-classification}
\end{figure}

Given a volumetric dataset, there is in general no default or ``natural'' way to obtain emission and absorption coefficients. Instead, the user must determine the visualization of the various structures present in the data, by assigning optical properties 
using an arbitrary (artificial) mapping, namely the transfer function.
While methods exist to automatically generate transfer functions~\cite{Kindlmann:1998:SAGTF}, the process of finding an 
appropriate transfer function is generally manual, tedious, and time-consuming. 
Arens and Domik (2010)~\cite{Arens:2010:START-TF} give a survey of transfer functions for volume rendering in which they classified transfer functions into the six categories: 1D data-based, gradient 2D~\cite{Levoy:1988:DSVD,Kniss:2002:MDTF}, curvature-based~\cite{Kindlmann:2003:CBTF}, size-based~\cite{Correa:2008:SBTF}, texture-based~\cite{Caban:2008:TBTF}, and distance-based~\cite{Tappenbeck:2006:DBTF}. 
A state-of-the-art (STAR) report on transfer functions for volume rendering has been published by Ljung et al.~\cite{Ljung:2016:START-TF}. This STAR report classified transfer function research based on the following aspects: dimensionality, derived attributes, aggregated attributes, rendering aspects, automation, and user interfaces (Figure~\ref{fig:TF-classification}).
Segmentation is the process of identifying individual voxels as belonging to one of the several materials. 
Rendering with segmented volume data~\cite{Rhee:2010:SBVA,Hadwiger:2003:HQT2LVR} can be seen as an extension of standard volume rendering, 
owing to the fact that different segmented regions can be rendered using different transfer functions.

\subsection{Volume Rendering of the Human Hand}
\label{sec:vr-hand}

\begin{figure}[tb]
 \centering % avoid the use of \begin{center}...\end{center} and use \centering instead (more compact)
 \includegraphics[width=1.0\hsize]{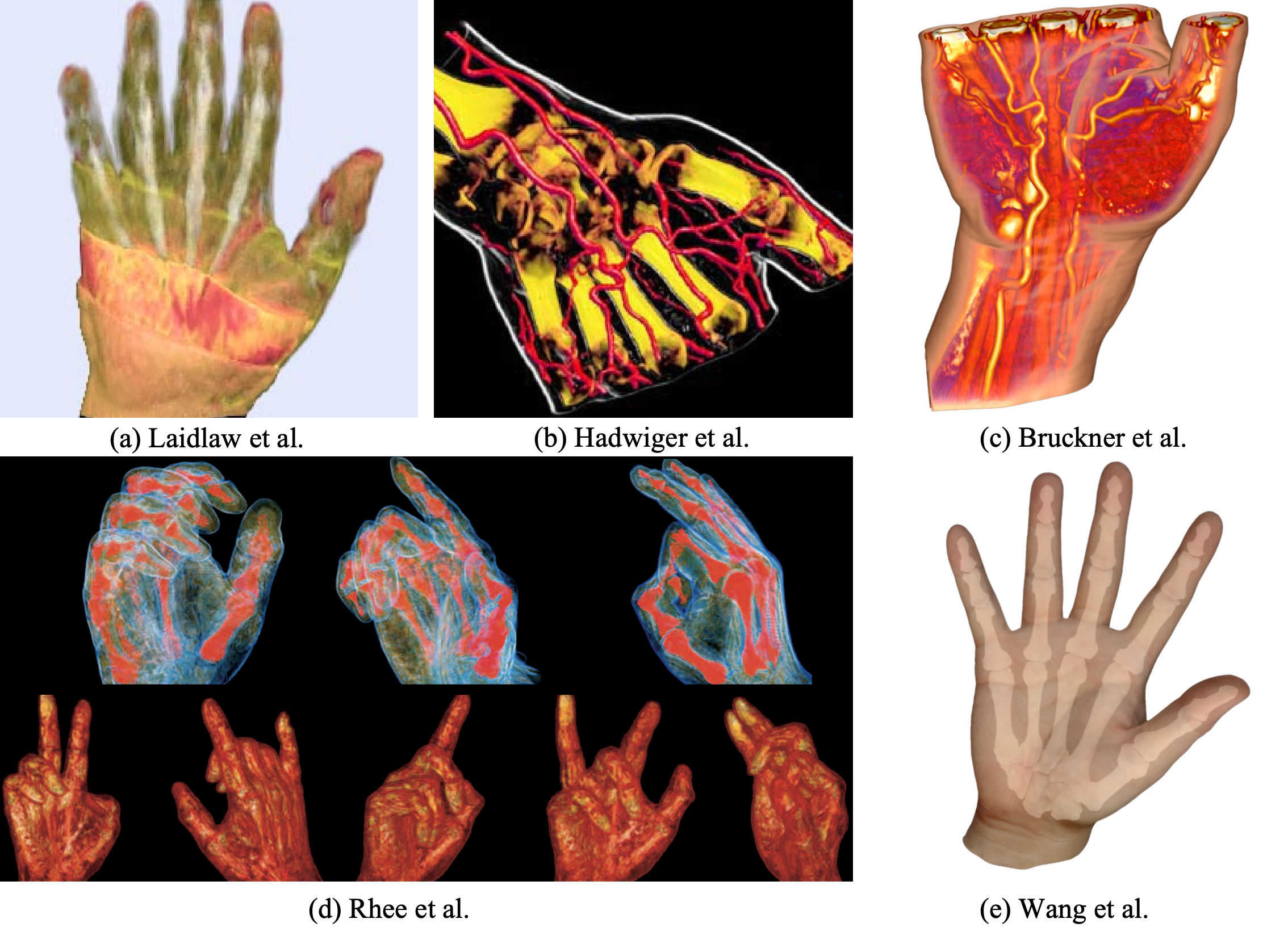}
 \caption{Examples of volume rendering from related work. Reprinted with permission.}
 \label{fig:hand-related-work}
\end{figure}

When volume-rendering medical scans, it is in general not possible to concurrently display all the data,
but instead, one typically visualizes selected parts or representations of the data.
Laidlaw et al. (1998)~\cite{Laidlaw:1998:PVB} (Figure~\ref{fig:hand-related-work}, (a)) allowed for a mixture of materials inside a voxel, which reduces segmentation artifacts. 
Hadwiger et al.~\cite{Hadwiger:2003:HQT2LVR} presented a two-level volume rendering approach using explicit segmentation information, where different objects can have different transfer functions and different rendering modes that are not only limited to DVR (Figure~\ref{fig:hand-related-work}, (b)).
Bruckner et al. (2006)~\cite{Bruckner:2006:ICPVR} (Figure~\ref{fig:hand-related-work}, (c)) proposed a context-preserving volume rendering model inspired by the technique of \textit{ghosting} from illustration. This is related to the concept of ``focus-plus-context'', 
which is well-known in information visualization: particularly interesting subsets of the data are considered to be ``in focus'', 
whereas the rest of the data merely provides context. 
Related to our work, Rhee et al.~\cite{Rhee:2010:SBVA} presented the first anatomically plausible 3D volume renders of the hand MRI data in motion, including bone animation and soft tissue deformation driven by a joint skeleton. The bones are segmented from the hand MRIs. In their work, two styles of renders were used. One uses a constant red color for bones and semi-transparent skin (Figure~\ref{fig:hand-related-work}(d), top); and the other uses a volume-rendering program developed by Kniss et al.~\cite{Kniss:2001:IVRMDT} (Figure~\ref{fig:hand-related-work}(d), bottom).
Wang et al.~\cite{Wang:2019:HMA} proposed a system to model and simulate the hand using MRI. They first segmented hand bone anatomy (meshes) in multiple poses using MRI, and interpolated and extrapolated them to the entire range of motion, essentially producing an accurate data-driven bone animation rig. Then, they simulated soft tissues using Finite Element Method (FEM), driven by the bone animation. They visualized their anatomy by compositing the bone and skin surface-rendered images using transparency (Figure~\ref{fig:hand-related-work}, (e)). 
They also publicly released their dataset~\cite{HandMRIDataset} of MRI scans of the human hand in multiple poses.
We use their dataset and their animated meshes in our work.

\section{Background}
\label{sec:theoretical background}

\subsection{Volume Ray Casting}
\label{sec:volumeRayCasting}

\begin{figure}
    \centering
    \includegraphics[width = 0.4\textwidth]{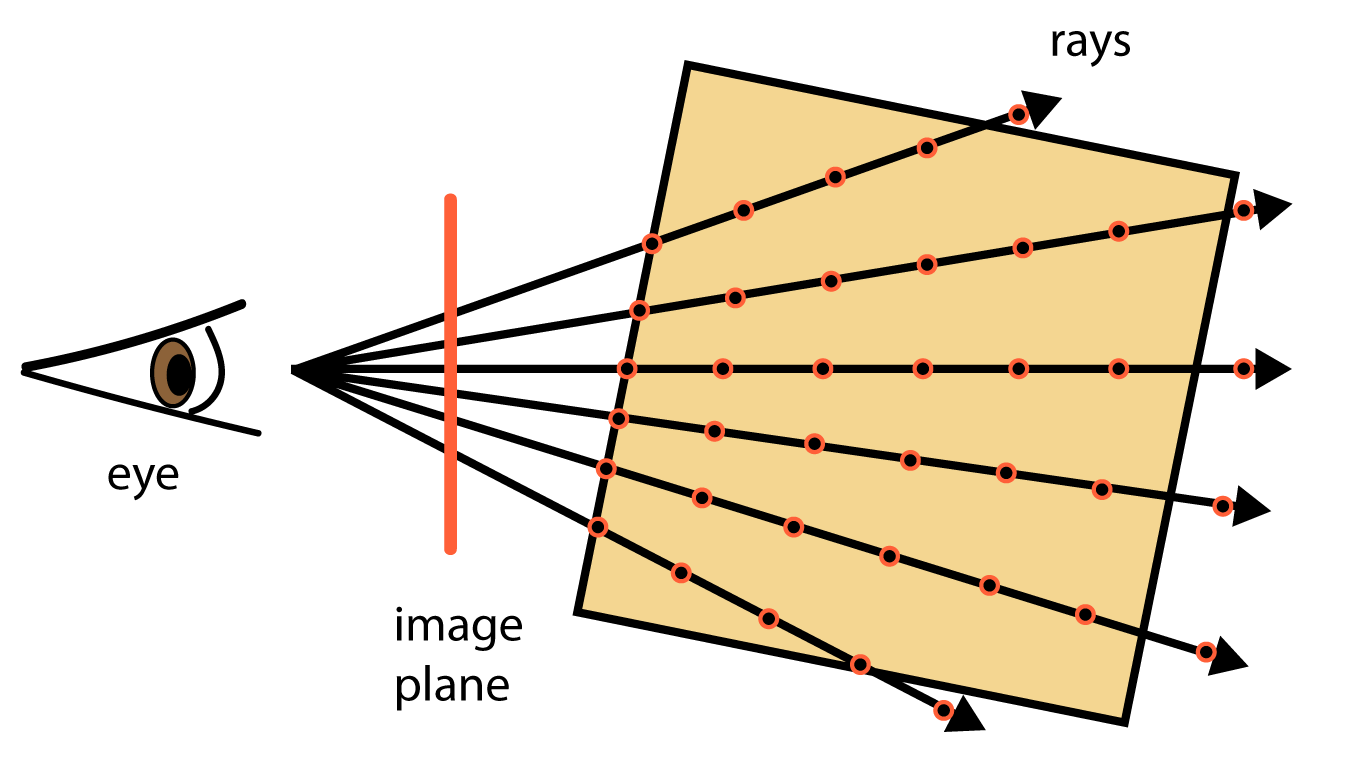}
    \caption[Ray-casting.]{\textbf{Ray-casting.} For each pixel, one ray is fired through the volume. 
     The ray is sampled at discrete positions to evaluate the volume-rendering integral.}
    \label{fig:basicRayCasting}
\end{figure}

We now describe volume ray casting~\cite{Kruger:2003:RAY-CASTING,Engel:2004:RTVG,Rezk:2008:AIT}
as used in our work; readers familiar with it can skip to Section~\ref{sec:handAnatomy}.
Volume rendering visualizes a 3D scalar field stored on a uniform grid in 3D.
To achieve this, scalar values are mapped to physical quantities ``emission color'' and ``opacity''
that describe how the light interacts with the volume at each 3D location (the \textit{transfer function}).
Generally, volume rendering includes emission, absorption and scattering~\cite{Max:1995:OMDVR}.
As is commonly done, we drop scattering and only retain and emission and absorption. This is
because this model provides a good compromise 
between output quality, computational efficiency, and complexity of tuning the transfer functions. 
Volume ray casting renders a 2D image by evaluating
the volume-rendering integral (Equation~\ref{eq:volumeRenderingIntegral}) along the camera rays
(Figure~\ref{fig:basicRayCasting}).
For each pixel in the rendered image, a single ray is cast into the volume. 
The volumetric scalar data is resampled from a discrete grid to (usually equispaced) points along the ray,
typically using tri-linear interpolation.
The resampled value is then mapped to optical properties using a transfer function, 
yielding an RGBA quadruplet whereby RGB is the emission color, and A is opacity (i.e., absorption).
The volume rendering integral is then approximated via either front-to-back or back-to-front 
compositing; as commonly done, we use the latter choice.
We denote a ray that has been cast into the volume by $\mathbf{x}(t)$, and parameterize it by the distance $t>0$ from the camera. 
The trilinearly interpolated scalar value at position $\mathbf{x}$ is denoted by $s(\mathbf{x}).$ 
Emissive color $c$ and absorption $\kappa$ are then functions of $t,$
\begin{equation}
    c(t) := c(s(\mathbf{x}(t)))\in\RR^3, \qquad \kappa(t) := \kappa(s(\mathbf{x}(t)))\in\RR.
\end{equation}
The pixel RGB color is computed by integrating 
emissive color $c(t)$ and absorption $\kappa(t)$ along each ray:
\begin{equation}
\label{eq:volumeRenderingIntegral}
  \textrm{pixelRGBColor} = \int_{t_\textrm{min}}^{t_\textrm{max}} c(t)e^{-\tau(t)} dt,\quad\textrm{for}\quad \tau(t) := \int_{t_\textrm{min}}^{t}\kappa(s) ds.
\end{equation}
Here, $\tau(t)\in\RR$ is the \emph{optical depth}, measuring net light absorption between the ray origin and $\mathbf{x}(t);$
smaller and larger values mean that the participating medium is more and less transparent, respectively.
The emissive color $c(t)$ has three components (R, G, B). Each component is treated separately.
The volume-rendering integral (Equation~\ref{eq:volumeRenderingIntegral}) typically
cannot be evaluated analytically, and numerical methods must be used instead.
Let $\Delta t$ denote the distance between successive sampling locations. 
It can be shown~\cite{Engel:2004:RTVG} that by defining 
the discrete opacity $\alpha$ and discrete emitted color $C$
\begin{gather}
\label{eq:discreteQuantities}
\alpha_i := 1 - e^{-\kappa(i\cdot\Delta t)\Delta t},\quad C_i = c(i\cdot \Delta t) \Delta t,\\
\textrm{(here,\ } i \textrm{\ is\ the\ index\ of\ the\ }i\textrm{-th\ ray\ segment)},
\end{gather}
one can approximate the volume rendering integral (Equation~\ref{eq:volumeRenderingIntegral}) 
using the recurrence
\begin{equation}
\label{eq:backToFront}
  \bar{C}_i = \alpha_i C_i + (1 - \alpha_i) \bar{C}_{i+1}.
\end{equation}
We start with 
$i_\textrm{max} = \left\lceil {t_\textrm{max}/\Delta t} \right\rceil,$ 
$\bar{C}_{i_\textrm{max}} = 0,$ 
and proceed down in back-to-front order to 
$i_\textrm{min} = \left\lfloor {t_\textrm{min}/\Delta t} \right\rfloor;$ 
the final volume rendering integral approximation (the pixel color) is $\bar{C}_{i_\textrm{min}}.$ 
The pixel calculations are independent and we used multi-threading to process pixels in parallel.
Theoretically speaking, the transfer functions map scalar values $s$ to 
$c$ and $\kappa;$ however, practically, we make them map to $C$ and $\alpha$ 
(defined in Equation~\ref{eq:discreteQuantities}).
This does make the $s\mapsto (c,\kappa)$ transfer functions dependent on $\Delta t,$ but in practice,
we set that parameter once and rarely tweaked it, and so this distinction was practically not
very significant for us.

\subsection{Human Hand Anatomy}
\label{sec:handAnatomy}

\begin{figure*}
    \centering
    \includegraphics[width = 1.0\textwidth]{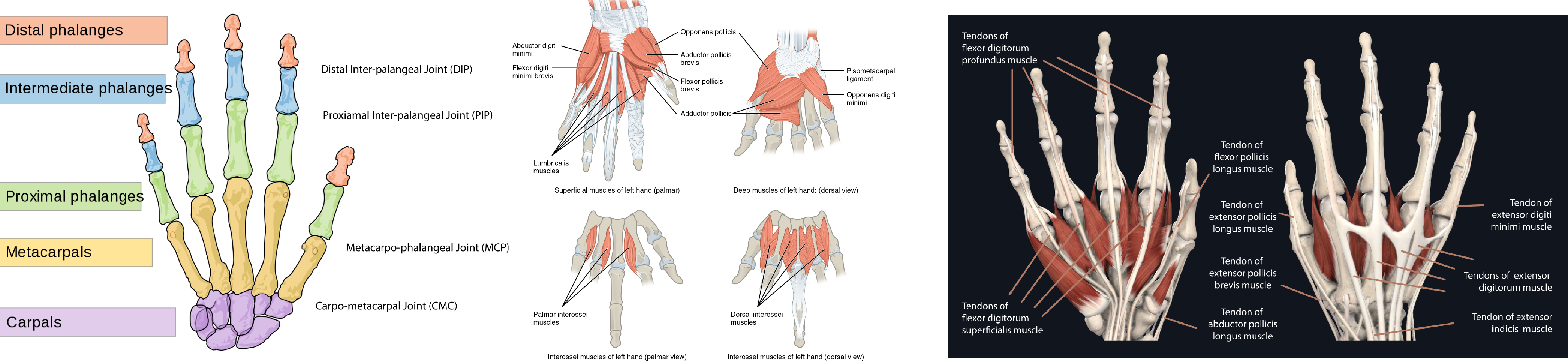}
    \caption{Left: \textbf{Bones and joints of the human hand} (Source: Wikimedia Commons). 
             Middle: \textbf{Muscles of the human hand} (Image downloaded from \url{https://commons.wikimedia.org/wiki/File:1121_Intrinsic_Muscles_of_the_Hand.jpg} by OpenStax under license CC-BY-4.0).
             Right: \textbf{Tendons of the human hand} ((C) 2022 Elsevier. Image adapted, with permission, from \url{https://3d4medical.com}).
            }
    \label{fig:anatomy}
\end{figure*}

The structure of the human hand is very complex. 
It is composed of multiple organs such as bones, joints, ligaments, muscles, tendons, blood vessels, nerves and skin~\cite{Hopkins,Arthritis,Cleveland}.
Bones are linked with joints. The hand has ligaments that hold the bones and cartilage together and provide flexibility. 
Muscles and tendons connect the bones, and through activation, create contractile forces and torques that bend the joints.
Figure~\ref{fig:anatomy} depict the bones, muscles and tendons, respectively.

\subsection{Acquisition of MRI data, segmentation and organ mesh simulation} 
\label{sec:dataAcquisition}

\begin{figure*}[tb]
 \centering % avoid the use of \begin{center}...\end{center} and use \centering instead (more compact)
 \includegraphics[width=\textwidth]{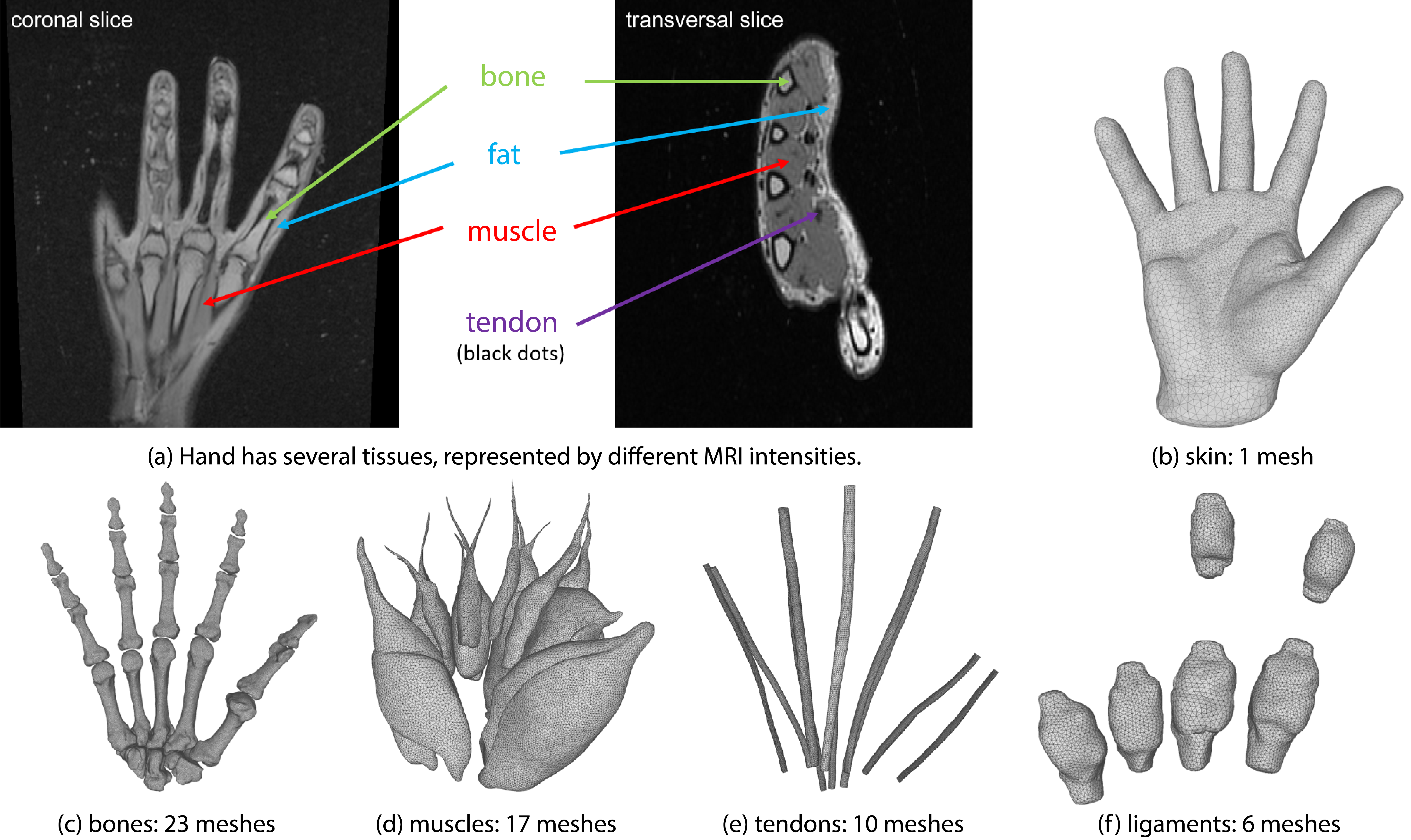}
 \caption{MRI slices (a) and corresponding mesh geometry (b-f) in the neutral pose.}
 \label{fig:mri-slice-and-geometry}
\end{figure*}

We obtained our hand MRI data (Figure~\ref{fig:mri-slice-and-geometry}(a)) 
from the project~\cite{HandMRIDataset}. The authors used lifecasting materials to generate 
hand molds to stabilize the hand during the scanning. 
Their scans contain a few air bubbles outside of the hand; but they are sufficiently
far away and not interfering with the hand volume.
In addition, we use their 3D segmentation of the MRI scans into triangle meshes of individual
musculoskeletal organs, obtained using the method described in~\cite{Wang:2019:HMA}.

The segmented organs are: skin, bones, muscles, tendons, and ligaments. The segmentation includes all hand bones, including phalanges, metacarpals, and carpals, with the exception of the proximal row of carpals; this is due to the low MRI quality at the wrist region (Figure~\ref{fig:mri-slice-and-geometry}(a), Figure~\ref{fig:mri-slice-and-geometry}(c)). All muscles of the hand were segmented (Figure~\ref{fig:mri-slice-and-geometry}(d)). All tendons (except extensor pollicis brevis and abductor pollicis longus located at the medial side of the thumb) were segmented (Figure~\ref{fig:mri-slice-and-geometry}(e)). Similarly, the ligaments (Figure~\ref{fig:mri-slice-and-geometry}(f)) at 4 MP joints of the index, middle, ring and pinky finger, and 2 PIP joints of the index and middle finger, were segmented and combined into six mesh ``capsules'' that wrap around these joints. The numbers of meshes for the different organs are: 1 for the skin, 23 for bones, 17 for muscles, 10 for tendons, and 6 for ligaments. The extracted geometry closely matches the MRI data: Figure~\ref{fig:mri-slice-and-geometry} shows the MRI slices and corresponding 
mesh geometry in the neutral pose. 

% Methods
\section{Volume Rendering of Human Hands}
\label{sec:pipeline}

We now explain how we apply volume rendering to human hands.
We assume that the segmentation of the MRI volume 
has already been performed, namely that the
meshes for the internal organs are known.
Our input consists of a grid-based MRI scalar volumetric
dataset, and the triangle meshes for the internal organs 
and the hand's external mesh (the skin). 
These meshes are allowed to overlap (due to segmentation errors), 
and some are contained inside other meshes. For example, all
internal organs are always contained inside the skin.
The project~\cite{HandMRIDataset} provided us with hand animations, namely
a temporally animated MRI, and temporally animated segmented organ meshes.
We can then apply volume-rendering separately to each frame, producing
volume-rendering videos of the anatomy in motion.
It is, however, conceptually important to consider just a single frame
when describing volume-rendering, as multiple frames are processed independently.

\begin{figure}
    \centering
    \includegraphics[width=0.4\textwidth]{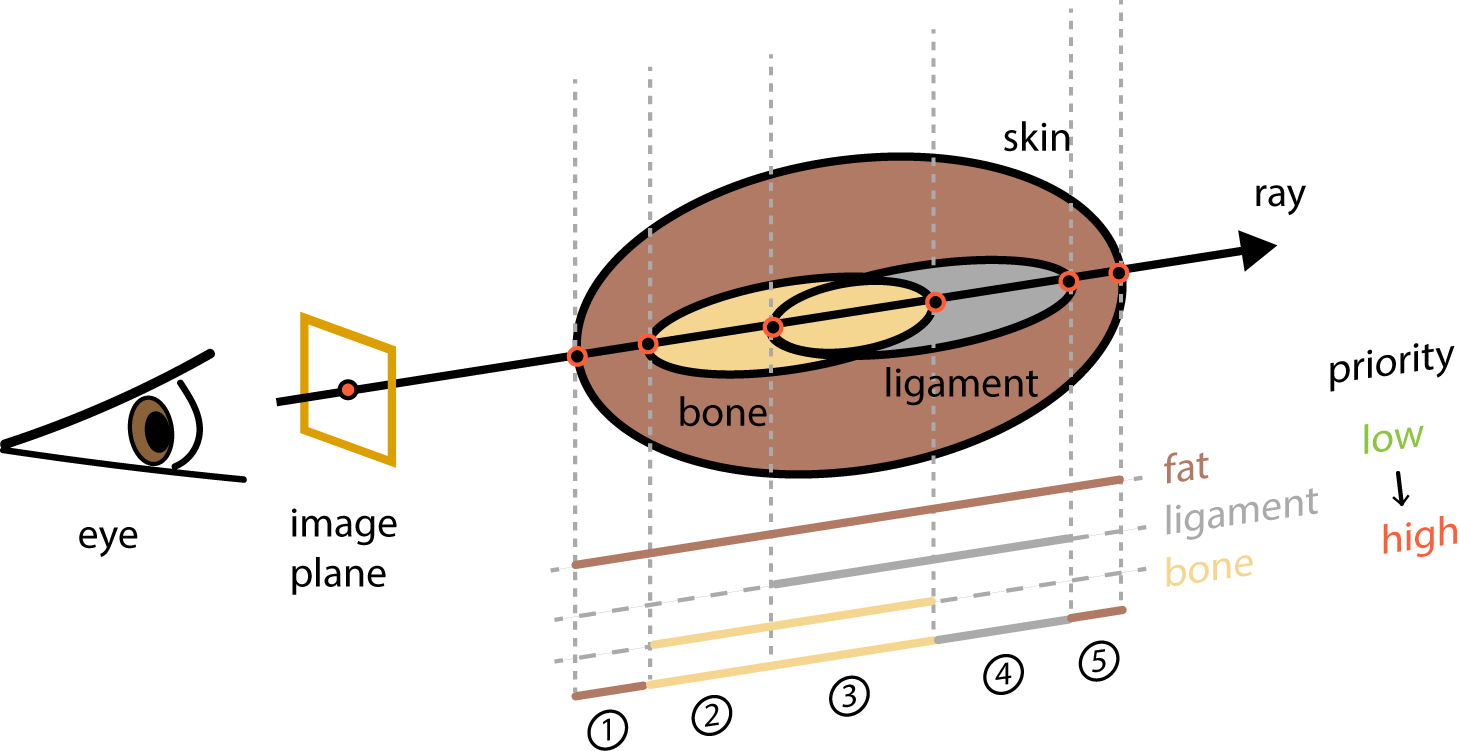}
    \caption[An example of material assignment.]
            {\textbf{An example of material assignment.} A single ray corresponding to a given image pixel is intersected with all meshes in the scene (skin, bone and ligament in this example). Sampling points that lie inside the interval between two neighboring intersections are identified as the same material. In the first and fifth intervals, only skin is present in the queue, and so every sampling point in these two segments is classified as fat tissue. In the second interval, both bone and skin are present in the queue. However, the bone has higher priority than the skin, so sample points in this region are classified as bone tissue. Similarly, sample points in the fourth interval are classified as ligament tissue. In the third interval, all tissues are present, hence the bone material is assigned because the bone has the highest priority.}
\label{fig:materialAssignment}
\end{figure}
\begin{figure}[tb]
 \centering % avoid the use of \begin{center}...\end{center} and use \centering instead (more compact)
 \includegraphics[width=1.0\hsize]{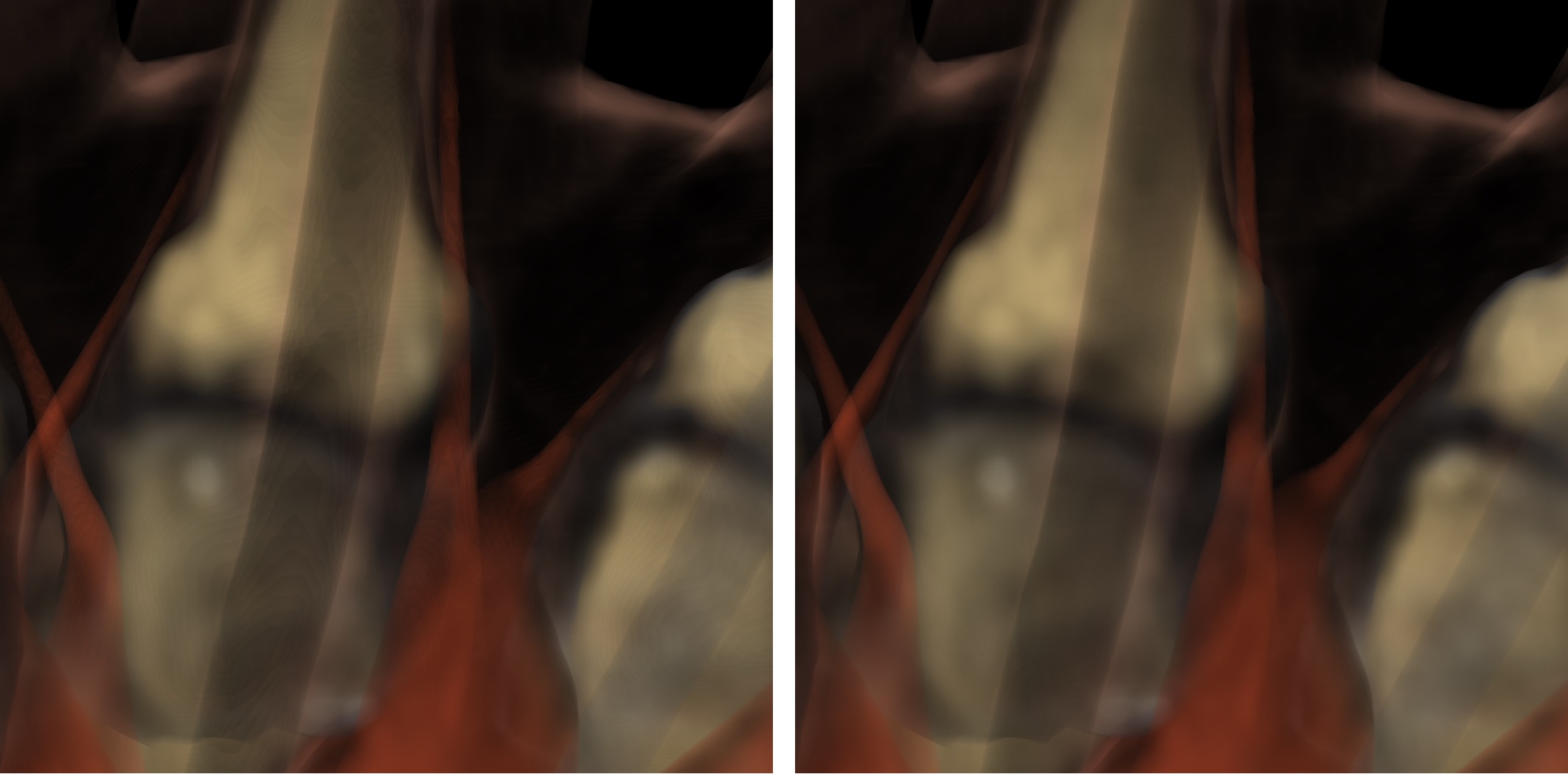}
 \caption[Volume rendering without and with stochastic jittering of the sampling positions.]{\textbf{Volume rendering without (left) and with (right) stochastic jittering of the sampling positions.} Wood-grain artifacts can be observed in the left image.}
 \label{fig:samplingArtifacts}
\end{figure}
For each pixel, a ray is cast into the MRI volume. 
We compute the intersections between the ray and the organ meshes, 
sorted by the distances to the camera. 
In this manner, each ray is decomposed into segments between 
consecutive intersections (Figure~\ref{fig:materialAssignment}).
Due to mesh intersections, the different segments may belong to one or multiple organs, and are assigned
individual transfer functions (Section~\ref{sec:transfer}). 
Because the empty space outside the hand does not contribute to the final image,
the first and last intersection define the starting and ending point of our sampling, respectively. 
This accelerates rendering, and effectively ignores any MRI noise outside the hand, improving image quality.
The volume integral is then approximated using Equation~\ref{eq:backToFront}, by evaluating
the MRI values at discrete samples along the ray, 
applying transfer functions, and compositing the resulting colors and 
opacities using back-to-front order (Section~\ref{sec:volumeRayCasting}). 
The MRI values are determined
by trilinearly interpolating the MRI values from the discrete grid points into the entire 3D volume.
We further improve image quality using stochastic jittering of the sample positions along the ray.
Namely, due to the discretization of Equation~\ref{eq:volumeRenderingIntegral} into Equation~\ref{eq:backToFront}, 
equidistant sampling leads to ``wood-grain artifacts'' (Figure~\ref{fig:samplingArtifacts}, left).
Figure~\ref{fig:samplingArtifacts} shows a comparison of two renders without and with stochastic jittering.
Under stochastic jittering, it is important to correct the opacity of the discretely sampled
points for use in Equation~\ref{eq:backToFront}. 
Namely, if opacity is $\alpha_{\textrm{equidistant}}$ under equidistant sampling $\Delta x_0,$ then
the corrected opacity under new sampling distance $\Delta x$ is
\begin{equation}
  \alpha_{\textrm{corrected}} = 1 - \bigl(1 - \alpha_{\textrm{equidistant}}\bigr)^{\Delta x / \Delta x_0}.
\end{equation}

\subsection{Material Assignment} 
\label{sec:material}

\begin{figure}
    \centering
    \includegraphics[width=0.5\textwidth]{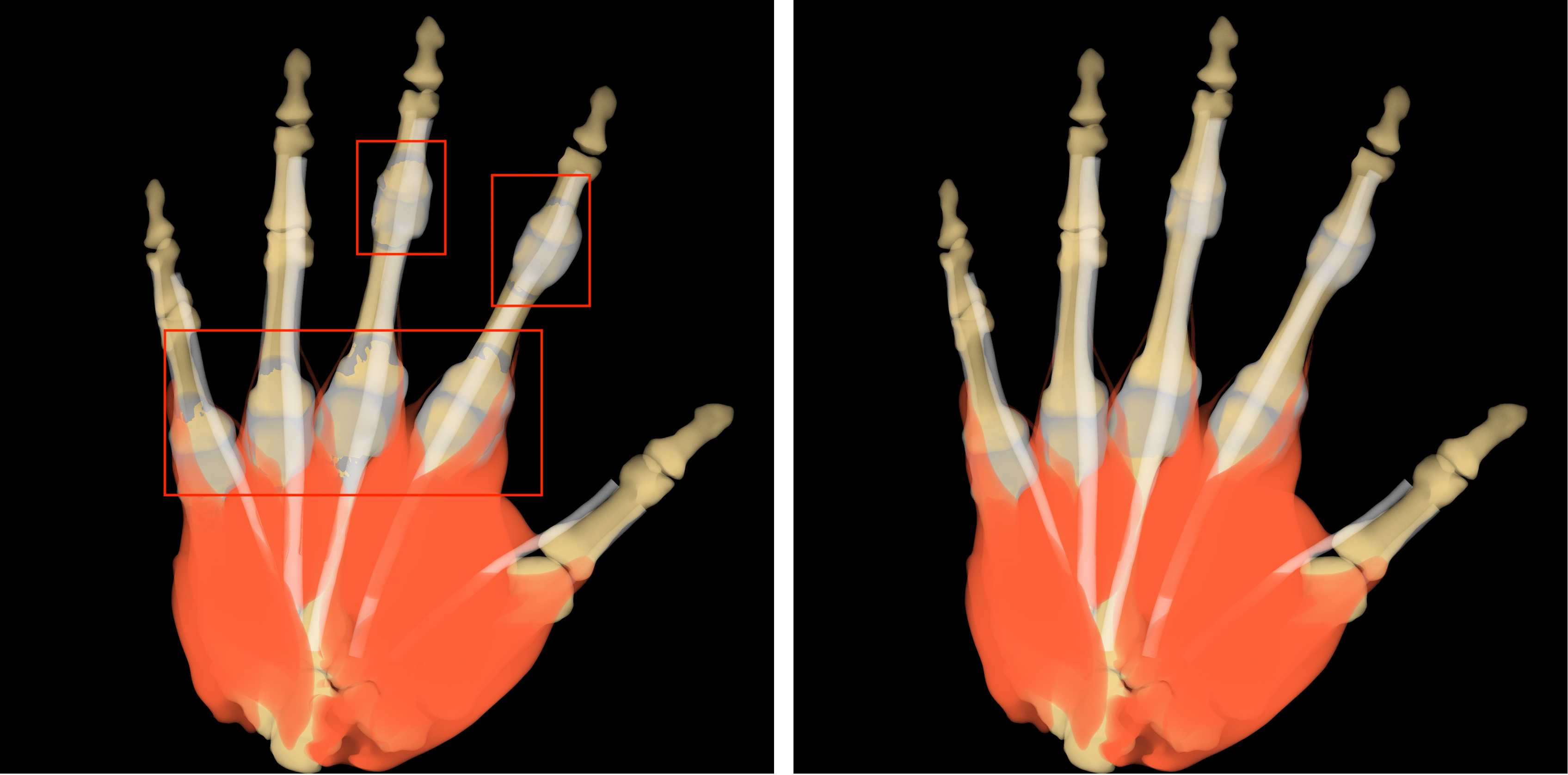}
     \caption[Comparison of renders without and with priority assignment.]{\textbf{Comparison of renders without (left) and with (right) priority assignment.} Artifacts marked with red squares are caused by wrong material assignments between bones and joint ligaments.
For better visibility, we used constant color and opacity as transfer functions (i.e., no MRI data).}
     \label{fig:priorityAssignment}
\end{figure}

\begin{figure}
    \centering
    \includegraphics[width=0.5\textwidth]{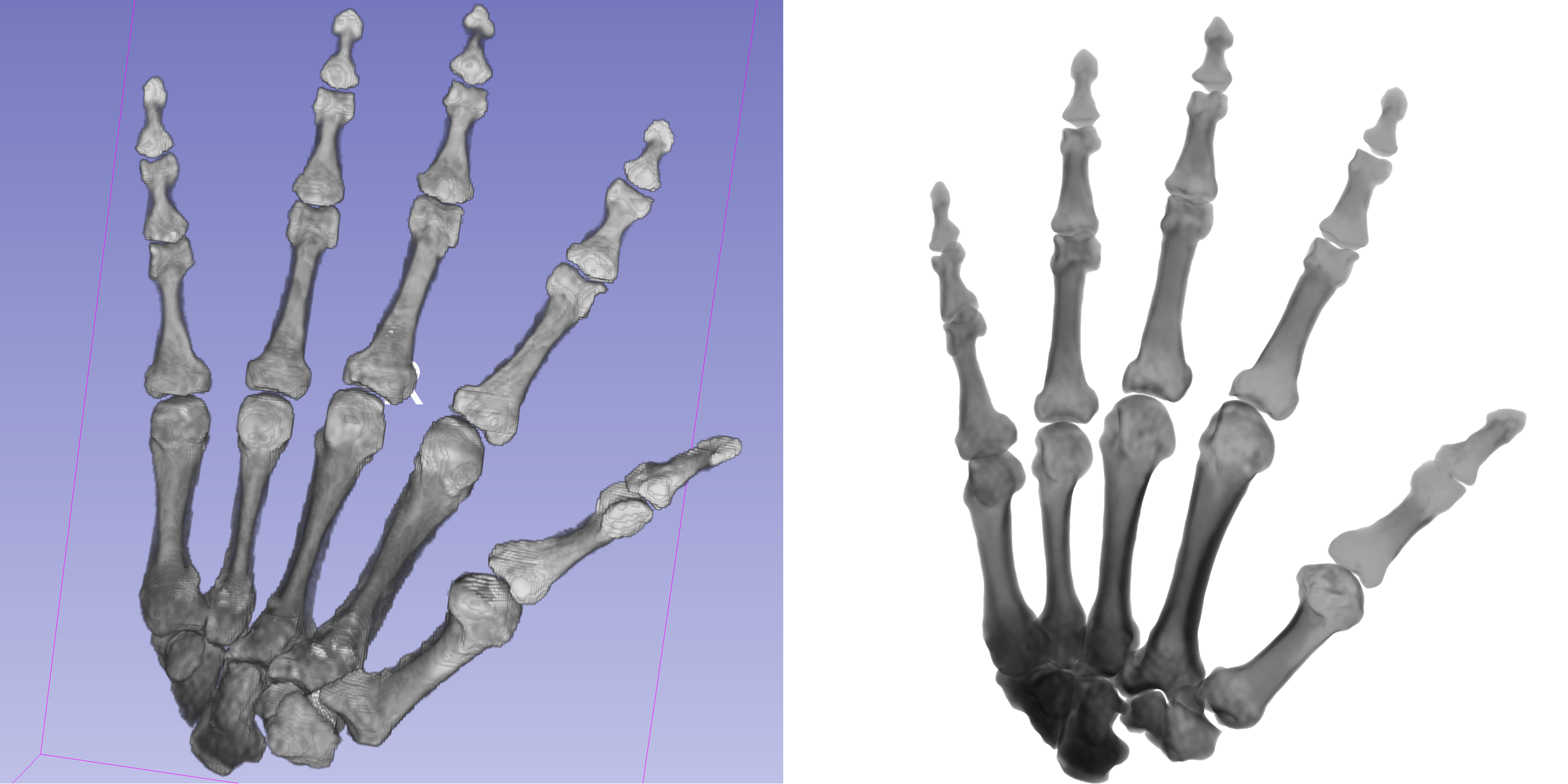}
     \caption[Comparison of segmented bone renders between ``3D slicer'' and our method.]{\textbf{Comparison of segmented bone renders between ``3D slicer'' (left) and our method (right).} Our result is free from staircase artifacts that can be observed on the left.}
    \label{fig:staircaseArtifacts}
\end{figure}

For each sample location $\mathbf{x}(t),$ we need to first determine which organ contains it,
so that we can then (in Section~\ref{sec:transfer}) use a correct transfer function for $\mathbf{x}(t).$
Organs are represented by meshes, and due to segmentation errors and simulation errors,
some of these meshes may overlap. For example, every organ mesh is inside the skin mesh, joint ligaments intersect with the end of the phalanges, etc.
Therefore, some samples may be inside the meshes of multiple organs, and we need a strategy to determine
a well-defined organ containing $\mathbf{x}(t).$ We do this by imposing the following priority rule:
\begin{equation}
\label{eq:materialPriority}
  \textrm{bone} > \textrm{tendon}  > \textrm{muscle} > \textrm{ligament} > \textrm{fat}.
\end{equation}
By definition, ``fat'' consists of locations that are inside the hand (i.e., in the space enclosed by the skin mesh),
but are not in any other organ.
Figure~\ref{fig:materialAssignment} gives an example of an material assignment.
Figure~\ref{fig:priorityAssignment} compares rendering without vs with priority assignment. 
We note that typically in volume rendering, segmentation into different material is performed at the level of each grid voxel.
This leads to ``staircase'' artifacts at the boundaries of two organs. Our approach of performing ray-mesh intersections and sampling within the segments between intersections avoids staircase artifacts, and produces higher image quality than the voxel-level-segmentation approach. Figure~\ref{fig:staircaseArtifacts} compares our method to ``3D slicer'', a free open-source software that uses voxel-level segmentation.

\subsection{Transfer Functions for Hand Anatomy}
\label{sec:transfer}

Because we know the material assignment at each sample, 
we can use different transfer functions for different organs (Section~\ref{sec:material}). Previously, 
Hadwiger et al. (2003)~\cite{Hadwiger:2003:HQT2LVR} and 
Bruckner et al. (2006)~\cite{Bruckner:2006:ICPVR} utilized focus-plus-context (F+C) to better visualize 3D datasets. 
Some objects were deemed (by the user) to be ``in focus'' and others not (the ``context'').
Objects ``in focus'' were rendered with a low or zero transparency, whereas 
the ``context'' objects served as a much more transparent reference. Inspired by these ideas, 
we devised two styles of transfer functions to emphasize either the inner organs (bones, muscles, tendons, and ligaments) 
or the subcutaneous fat tissue beneath the skin. We call these two 
styles ``interior-emphasized'' and ``fat-emphasized.'' 
A style is specified by giving a transfer function for each material. 

\subsubsection{Interior-Emphasized Style}

\begin{figure}
    \centering
     \includegraphics[width=0.5\textwidth]{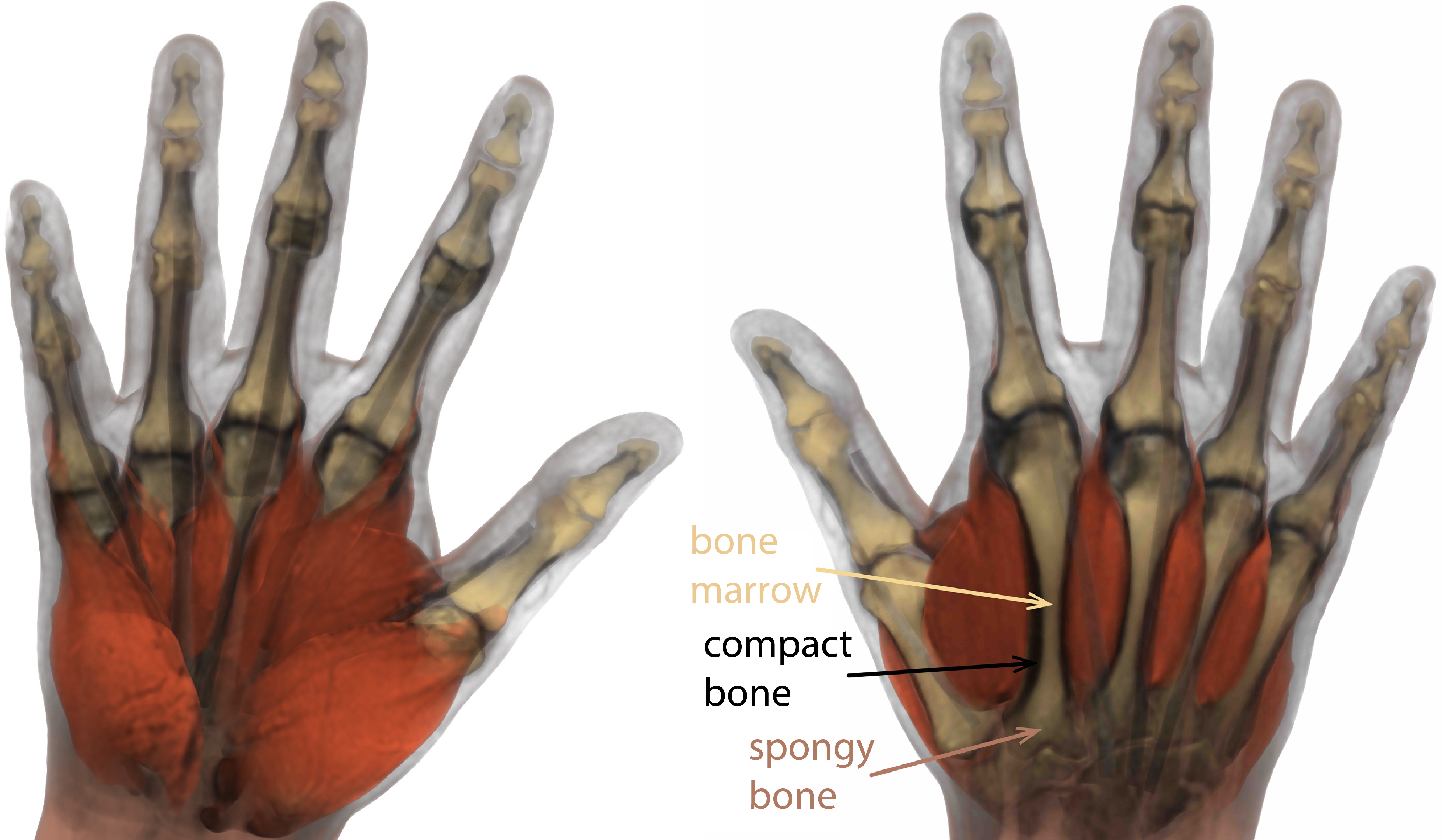}
    \caption[Volume rendering with interior-emphasized style.]{\textbf{Volume rendering with interior-emphasized style.} Internal organs are clearly shown (on both sides), boundaries between tissues are smooth and clear, and semi-transparent skin provides the context of the hand shape. In particular, on the palmar side (left), rich muscle textures can be observed; on the dorsal side (right), bone marrow, compact bone, and spongy bone of metacarpals can be recognized.}
    \label{fig:bone-style}
\end{figure}
In this style, the fat is visualized as a semi-transparent membrane, 
whereas the internal organs (bones, muscles, etc.) are visualized in a clear manner 
with high opacity and contrasting colors (Figure~\ref{fig:bone-style}). 
\begin{table}
\begin{adjustbox}{width=1.0\hsize, center}
% \centering
\begin{tabular}{c | c |cc}
\toprule
Material &  \text{Transfer Function} & \text{$C_{\textrm{material}}$}  & \textbf{$\alpha_{\textrm{material}}$} \\
\midrule
bone   &    & \text{(244, 214, 145)} \cellcolor{bone} &  1.0 \\
muscle  &  \text{$C = \max(\min(a\bigl(s/s_{\textrm{max}}\bigr)^{b}, 1.0), 0.0) * C_{\textrm{material}}$, } & \text{(255, 98, 56)} \cellcolor{muscle} & 1.0          \\
ligament&  \text{$\alpha = \alpha_{\textrm{material}}$.} & \text{(170, 170, 170)} \cellcolor{ligament}  & 1.0       \\
tendon  &     & \text{(255, 255, 255)} \cellcolor{tendon} & 1.0        \\
\midrule
\multirow{2}*{fat} & \text{$C = \rho_{\textrm{fat}}(s)/\rho_{\textrm{fat\_max}} * C_{\textrm{material}}$,} & \cellcolor{skin} & \multirow{2}*{0.6} \\
 & \text{$\alpha = \rho_{\textrm{fat}}(s)/\rho_{\textrm{fat\_max}} * \alpha_{\textrm{material}}$.} & \multirow{-2}*{\text{(177, 122, 101)}}\cellcolor{skin} & \\
\bottomrule
\end{tabular}
\end{adjustbox}
\caption[Transfer functions for interior-emphasized style.]{\textbf{Transfer functions for interior-emphasized style.} 
$s$ denotes the MRI scalar value; $s_{\textrm{max}}$ denotes the maximum MRI value in the dataset; 
$\rho_{\textrm{fat}}(s)$ denotes the frequency of MRI value $s$ within fat region; 
$\rho_{\textrm{fat\_max}}$ denotes the maximum frequency of MRI values in the fat region. 
All values were carefully chosen for best visualization.}
\label{tbl:interiorEmphasized}
\end{table}
In this style, organs are divided into two groups: non-fat tissues (bones, muscles, ligaments, tendons) and the fat tissue. 
For non-fat tissues, the color is the product of a tissue-specific constant color, and a scale factor determined from
the MRI value (see Table~\ref{tbl:interiorEmphasized}), clamped to $[0,1].$ 
The reason for using the scale factor that incorporates the MRI value is to expose the
anatomical detail, as these organs are ``in focus''.
Opacity is constant per non-fat tissue.
The fat tissue color is a constant color, scaled with a normalized frequency of this MRI value.
Namely, we build a histogram of all fat MRI values, and the height of a bin (the ``frequency'') 
$\rho_{\textrm{fat}}(s)$ holding the MRI value $s$ is used in the transfer function, normalized
by the maximum height of a bin $\rho_{\textrm{fat\_max}}(s)$ across the entire fat (Table~\ref{tbl:interiorEmphasized}).
\begin{figure}
    \centering
    \includegraphics[width=0.4\textwidth]{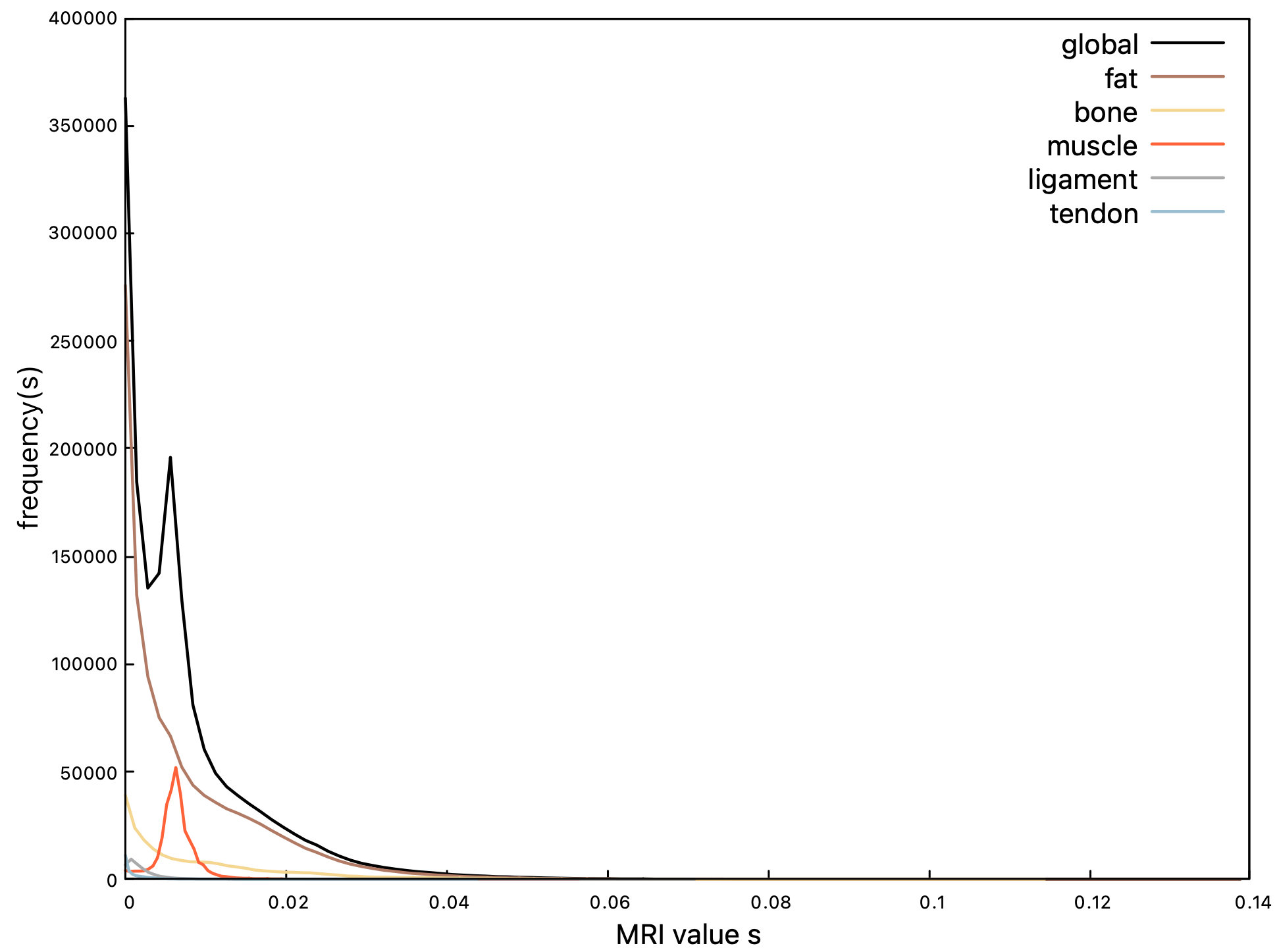}
    \caption{Frequency histogram of MRI values in the neutral pose.}
    \label{fig:frequencyHistogram}
\end{figure}
The reasons to use the frequency of MRI values for fat are twofold. First, in the past, histogram analysis has been utilized to extract tissue features~\cite{Laidlaw:1998:PVB,Lundstrom:2006:LH}; it is intuitive to highlight the data in regions corresponding to histogram peaks. Second, in particular for fat tissue, we found that in the wrist region, the MRI signal has lower quality than elsewhere (Figure~\ref{fig:mri-slice-and-geometry}). This introduces a substantial amount of low MRI values around the wrist. The MRI values of the skin happen to be in the same range as the wrist. Therefore, by utilizing the frequency of MRI values, we highlight the skin and partially hide the subcutaneous fat tissue, which has higher MRI intensity. Note that low intensity voxels in the wrist area are also emphasized, but they do not occlude the internal organs. Figure~\ref{fig:frequencyHistogram} shows the frequency histogram of the MRI value in all regions.

\subsubsection{Fat-Emphasized Style}

\begin{figure}
    \centering
    \includegraphics[width=0.5\textwidth]{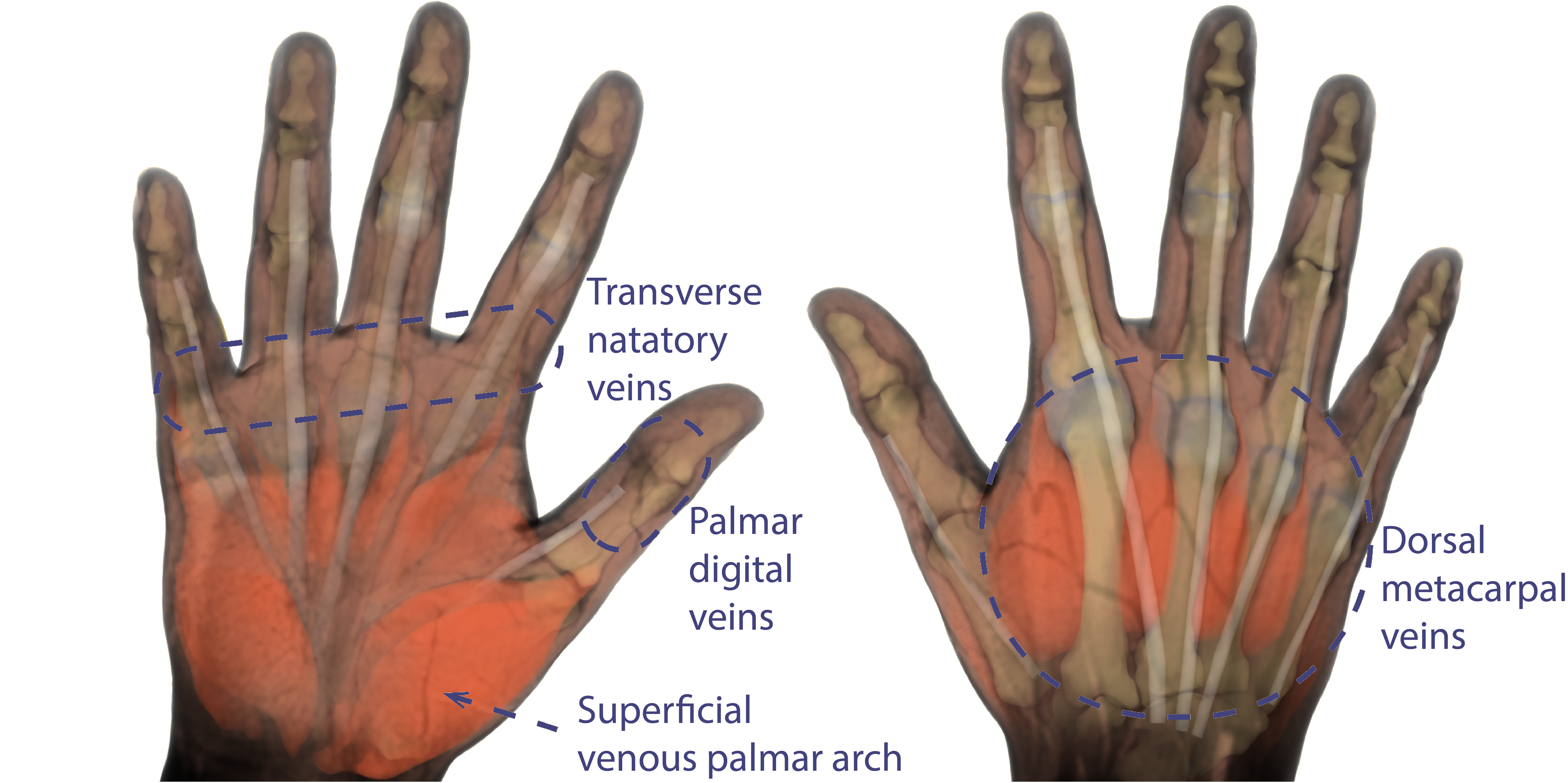}
    \caption[Volume rendering with fat-emphasized style.]
            {\textbf{Volume rendering with fat-emphasized style.} Fat layer is ``in focus'' and has significant opacity, while internal organs provide the ``context'' of shape and spatial relations. Several superficial veins underneath the skin are visible and annotated, such as the traverse natatory veins, palmar digital veins, superficial venous palmar arch, and dorsal metacarpal veins.}
    \label{fig:vein-style}
\end{figure}
In this style, the subcutaneous fat tissue is ``in focus'' and displayed with high opacity. Other tissues have constant but contrasting colors, which provides the context of shape and spatial relation. This style highlights veins in the fat layer, such as the traverse natatory veins, palmar digital veins, the superficial venous palmar arch, and dorsal metacarpal vein (Figure~\ref{fig:vein-style}). 
Vein visualizations are more obvious on the dorsal side because the dorsal veins are closer to the skin. 
In this style, each non-fat tissue is assigned a per-tissue constant color and opacity. 
Similar to the non-fat tissues in the ``interior-emphasized'' style, the fat tissue color is a constant color modulated with the MRI value to expose the anatomical detail, and the fat opacity is constant (Table~\ref{tbl:fatEmphasized}). 
The veins are visible under our ``fat-emphasized'' style because they have a lower MRI intensity than other tissues in the fat layer. 
It is important to note that our MRI scans used no contrasting media such as those utilized in magnetic resonance angiography (MRA), 
nor did we segment the veins. 
\begin{table}
\begin{adjustbox}{width=1.0\hsize, center}
% \centering
\begin{tabular}{c | c | cc}
\toprule
Material &  \text{Transfer Function} & \text{$C_{\textrm{material}}$}  & \textbf{$\alpha_{\textrm{material}}$} \\
\midrule
bone   &    & \text{(244, 214, 145)}\cellcolor{bone} &  1.0 \\
muscle  &  \text{$C = C_{\textrm{material}}$, } & \text{(255, 98, 56)}\cellcolor{muscle}  & 1.0          \\
ligament&  \text{$\alpha = \alpha_{\textrm{material}}$.} & \text{(170, 170, 170)}\cellcolor{ligament}  & 1.0       \\
tendon  &     & \text{(255, 255, 255)}\cellcolor{tendon} & 1.0        \\
\midrule
\multirow{2}*{fat} & \text{$C = \max(\min(a(s/s_{\textrm{max}})^{b}, 1.0), 0.0) * C_{\textrm{material}}$,} & \cellcolor{skin} & \multirow{2}*{0.6} \\
 & \text{$\alpha = \alpha_{\textrm{material}}$.} & \multirow{-2}*{\text{(177, 122, 101)}}\cellcolor{skin} & \\
\bottomrule
\end{tabular}
\end{adjustbox}
\caption[Transfer functions for fat-emphasized style.]
{\textbf{Transfer functions for the fat-emphasized style.} $s$ denotes the MRI scalar value, 
and $s_{\textrm{max}}$ denotes the maximum MRI value in the dataset. 
All values were carefully chosen to best visualize the data.}
\label{tbl:fatEmphasized}
\end{table}

% Results
\section{Results}
\label{sec:results}

We evaluate our volume rendering method using animated sequences of MRIs. 
We compare the results with standard surface rendering, and with previous volume rendering methods.

\subsection{Computational and Memory Performance}

Our CPU volume renderer was written in C++ and built and tested on Ubuntu 20.04 Linux, 
on an Intel Core i7-7700K processor with 8 cores and 40 GB of RAM (2400 MHz DDR4).
We used Intel\textsuperscript{®} Embree library (version 3.4.0) for calculating ray-object intersections 
and oneTBB (version 2021.5.1) for computing multiple pixels in parallel. 
Python scripts were used to automate batch rendering of testing sequences.
Our animated hand MRI data and mesh geometry for testing was obtained from the project~\cite{Wang:2019:HMA,HandMRIDataset}. 
These animated MRIs come from layered FEM simulation driven by five input joint animations: "close the fist" (keyframed, 132 frames), "opposition of the thumb" (keyframed, 996 frames), "performance animation" (motion-captured, 653 frames), "numbers 1-5" (motion-captured, 360 frames) and "American Sign Language" (keyframed, 732 frames). 

The resolution of the input MRI volume data is $400\times 400 \times400$ and the voxel size is $0.64mm \times 0.64 mm \times 0.64mm$. Table~\ref{tab:meshSpec} shows the specifications of mesh geometry used for classification. We rendered these five motion sequences with the two types of transfer functions (fat-emphasized, interior-emphasized) with three camera configurations (front, back, side), thus generating $5\times2\times3=30$ image sequences in total. On average, it took 3.7 seconds and 4.7 seconds to render an image of $1024 \times 1024$ resolution with fat-emphasized and interior-emphasized style of transfer functions, respectively. Table~\ref{tab:time} gives the time costs. Maximum memory usage was 612.3 MB.

\begin{table}
    \begin{adjustbox}{width=1.0\hsize, center}
    \begin{tabular}{l|ccc}
    \toprule
    tissue type & \# meshes & \# vertices & \# triangles \\
    \midrule
    bones &  23 & 788/8,291/3,353/77,111 & 1,572/17,838/6,764/15,576    \\
    muscles &  17 & 1,089/10,924/3,247/55,205 & 2,174/21,844/6,442/109,506 \\
    tendons & 10 & 3,000/9,032/7,613/76,129 & 5,996/18,060/15,222/152,218  \\
    ligaments & 6 & 663/1,658/1,051/6,305 & 1,322/3,312/2,098/12,586  \\
    skin & 1 & 18,105 & 36,206 \\
    \bottomrule
    \end{tabular}
    \end{adjustbox}
    \caption[Specifications of meshes of all hand tissues.]
            {\textbf{Geometric complexity of the tissue meshes.} Column "\# meshes" shows the number of meshes. Columns "\# vertices" and "\# triangles" give the minimum/maximum/average/total numbers of vertices and triangles for each tissue type.}
    \label{tab:meshSpec}
\end{table}

\begin{table}
    \begin{adjustbox}{width=1.0\hsize, center}
    \begin{tabular}{l|ccc|ccc|cc}
    \toprule
    \multirow{2}{5em}{sequences} &  \multicolumn{3}{c|}{fat-emphasized} & \multicolumn{3}{c|}{interior-emphasized} & \multirow{2}{1.5em}{\centering $n_\textrm{f}$} & \multirow{2}{3em}{\centering$t$ (hr)}  \\
     & $t_{\textrm{ff}}$ (s) & $t_{\textrm{fb}}$ (s) & $t_{\textrm{fs}}$ (s) & $t_{\textrm{if}}$ (s) & $t_{\textrm{ib}}$ (s) & $t_{\textrm{is}}$ (s) & & \\
    \midrule
    "Close the fist" & 3.60 & 3.54 & 3.45 & 4.68 & 5.24 & 4.05 & 132 & 0.9 \\
    "Opposition of the thumb" & 3.87 & 3.82 & 3.76  & 4.81 & 5.04 & 4.49 & 996 & 7.1 \\
    "Performance animation"  & 3.65 & 3.66 & 3.50 & 4.52 & 5.14 & 4.14 & 653 & 4.5 \\
    "Numbers 1-5" & 3.79 &  3.79 & 3.63 & 4.68  & 5.12 & 4.54 & 360  & 2.6 \\
    "American Sign Language" & 3.67 & 3.64 & 3.50 & 4.49 & 4.85 & 4.44 & 732 & 5.0 \\
    \bottomrule
    \end{tabular}
    \end{adjustbox}
    \caption[Rendering time cost for each animation sequence.]
            {\textbf{Rendering time cost for each animation sequence.} 
     The time $t_{xy}$ refers to ``fat-emphasized'' and ``interior-emphasized'' style for $x=\textrm{f},\textrm{i},$ respectively;
     and to front view, back view and side view for $y=\textrm{f},\textrm{b},\textrm{s},$ respectively.
     Columns $n_\textrm{f}$ and $t$ give the total number of animation frames, and total render time for each sequence, respectively.}
    \label{tab:time}
\end{table}

\subsection{Comparisons}

\begin{figure}
    \includegraphics[width=0.5\textwidth]{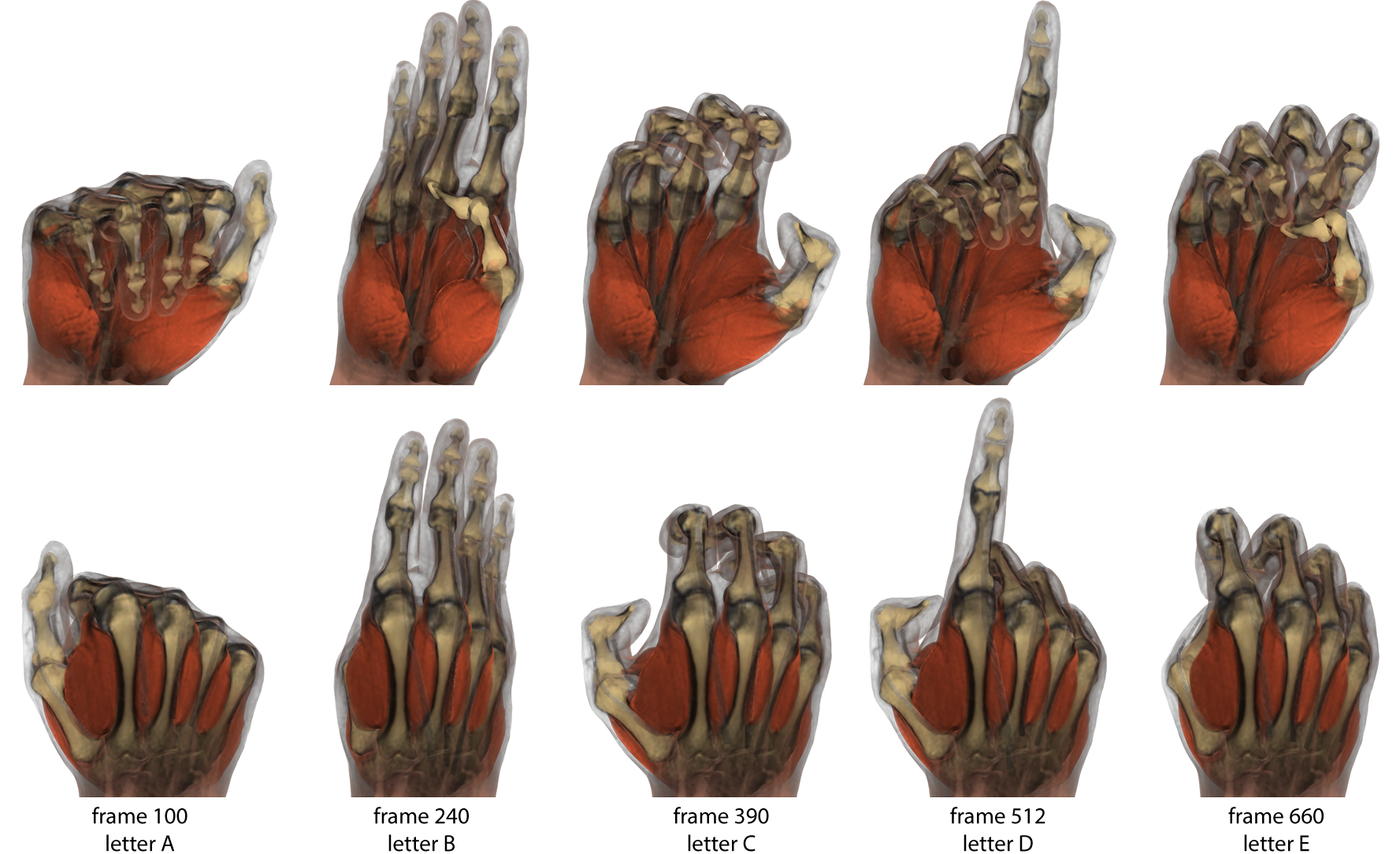}
    \caption[``American Sign Language'': letters A to E, rendered with ``interior-emphasized'' style; front and back view.]
    {\textbf{``American Sign Language''}: Letters A to E, rendered with ``interior-emphasized'' style; front view (top row) and back view (bottom row). The internal structures are clearly shown, the boundaries between tissues are smooth and clear, and the semi-transparent skin provides the ``context'' of the hand shape.}
    \label{fig:asl}
\end{figure}

In Figure~\ref{fig:asl}, we show the volume rendering of the sequence ``American Sign Language'' with the interior-emphasized style. The hand was posed into letters A-E and rendered from both front and back views. In these renderings, color and opacity are consistent across the range of motion. The internal structures are clearly shown, the boundaries between tissues are smooth and clear, and the semi-transparent skin provides the ``context'' of the hand shape. 
Due to using MRI values in their transfer function, muscles are visible with a rich anatomical ``texture'' on the palmar side, 
and the anatomy of metacarpals can be easily identified on the dorsal side.
The outer bone layer (compact bone), bone center (bone marrow)
and the two bone ends are shown in black, yellow and dark yellow, respectively.

\begin{figure}
    \includegraphics[width=0.5\textwidth]{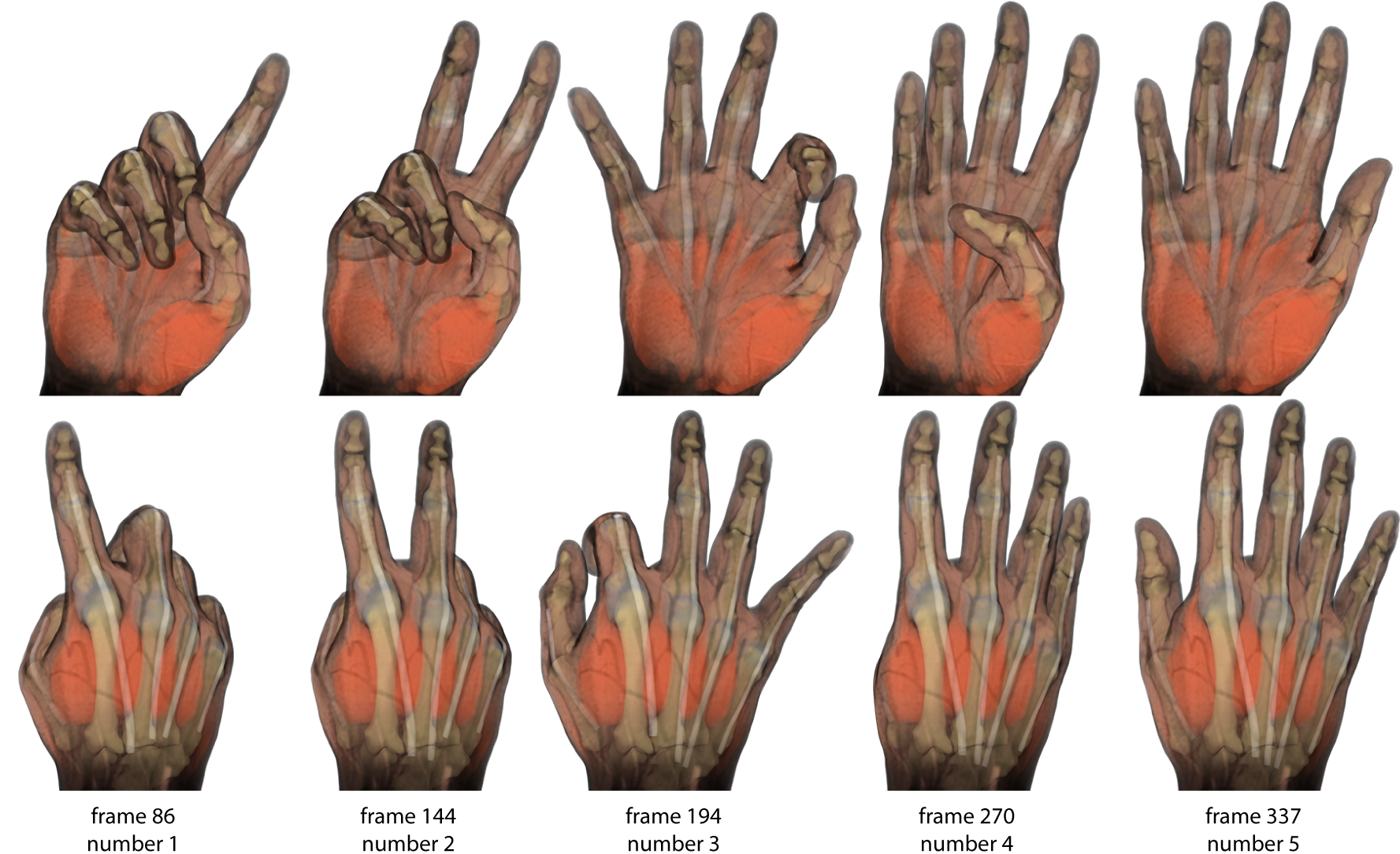}
    \caption[``Numbers 1-5'', rendered with the ``fat-emphasized'' style; front and back view.]
        {\textbf{``Numbers 1-5''}, rendered with fat-emphasized style, front view (top) and back view (bottom).}
    \label{fig:chinese1-5}
\end{figure}

In Figure~\ref{fig:chinese1-5}, we show the sequence ``numbers 1-5'' with the fat-emphasized style, 
in which the hand was posed into numbers 1 to 5, and rendered from front and back views. 
Hand and internal organs appearance is consistent across the range of motion. 
Constant but contrasting colors of the internal tissues provide shape cues for all the inner structures. 
The superficial veins beneath the skin are clearly visible (black color), 
and their motion is evident in the animation across the range of motion. 
Visualization of the interior anatomy are also useful to ``debug'' FEM simulations,
as any FEM simulation instabilities in the fat layer are much easier to identify in the ``fat-emphasized'' style,
compared to the surface rendering of the skin.

\begin{figure}
    \centering
    \includegraphics[width=0.5\textwidth]{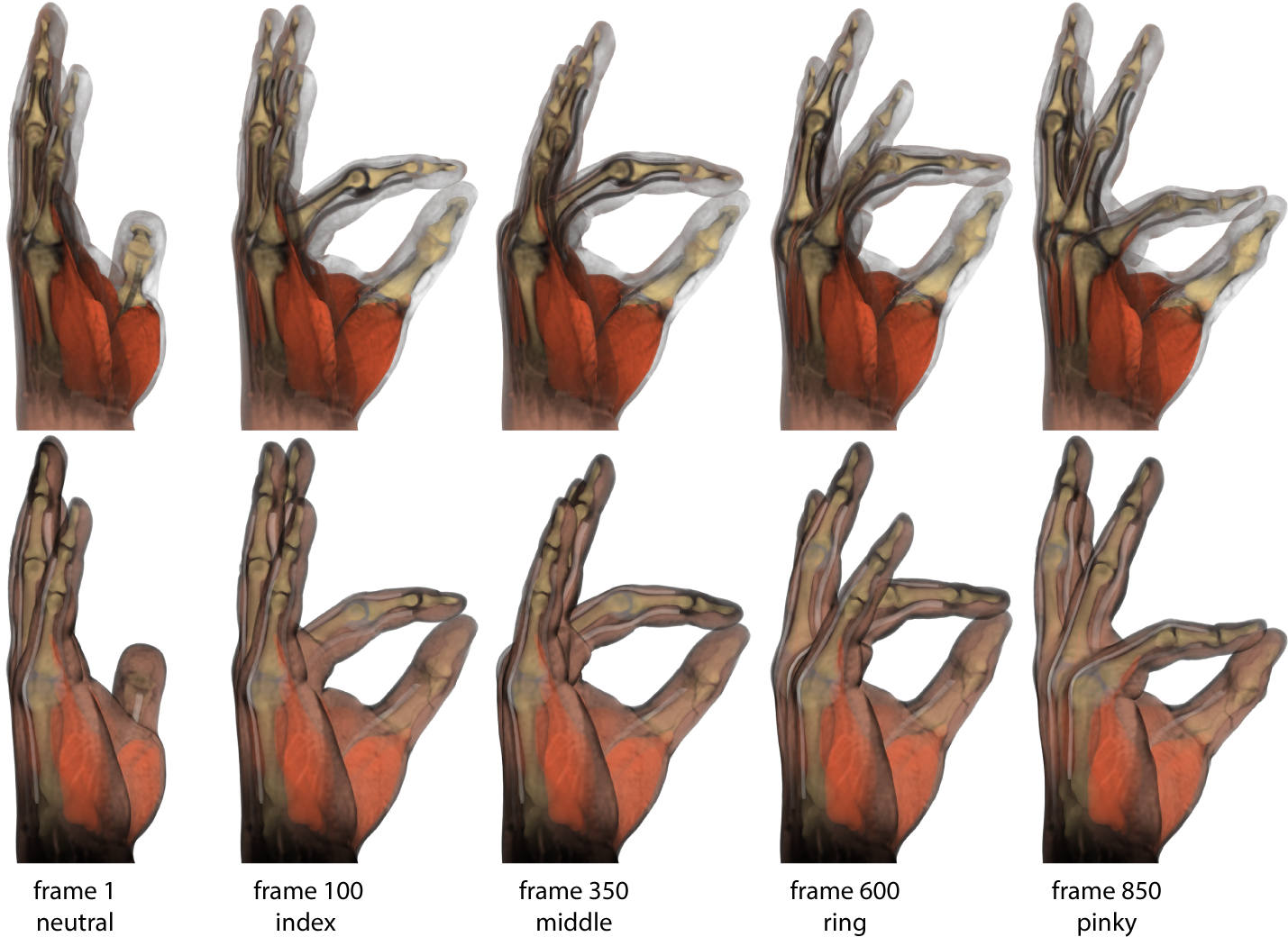}
    \caption[``Opposition of the thumb,'' rendered from the side view with interior-emphasized and fat-emphasized styles.]{\textbf{``Opposition of the thumb''}, rendered from the side view in the interior-emphasized (top) and fat-emphasized (bottom) styles.}
    \label{fig:opposition}
\end{figure}

Figure~\ref{fig:opposition} gives a comparison of ``fat-emphasized'' and ``interior-emphasized'' styles, 
using the sequence ``opposition of the thumb'' (side view). 
It is important to note that the two styles are designed to emphasize different anatomical structures 
while preserving the overall context, and not to compete with each other. 
The key visual differences between the two styles are: (1) bone anatomy is clearly visible in the interior-emphasized style, especially in middle and proximal phalanges and metacarpals, whereas the bone tissue in the fat-emphasized style merely provides the shape of its outer surface; (2) shapes and spatial positioning of tendons and ligaments are visible in the interior-emphasized style, but are not as easily observable as in the fat-emphasized style, due to low MRI values; (3) in the interior-emphasized style, skin and wrist regions are substantially visible, and the subcutaneous fat tissue is partially hidden, whereas in the fat-emphasized style, subcutaneous fat tissue is substantially visible, and skin and wrist region are close to black; (4) superficial veins more easily identifiable in the fat-emphasized style. 
The different camera views provide insight into different aspects of the hand. Specifically, in the interior-emphasized style, the front view emphasizes palmar muscles and phalanges due to the spatial proximity of muscles and bones. In the back view, the metacarpals and carpals that were occluded in the front view, are now visible. In addition, the side view provides insights into the spatial relationships between tendons, ligaments, and interphalangeal joints. Similarly, the fat-emphasized style not only provides shape cues for the above-mentioned anatomy, but also visualizes the anatomical MRI detail and superficial veins within the fat tissue. 

\begin{figure}
    \centering
    \includegraphics[width=0.5\textwidth]{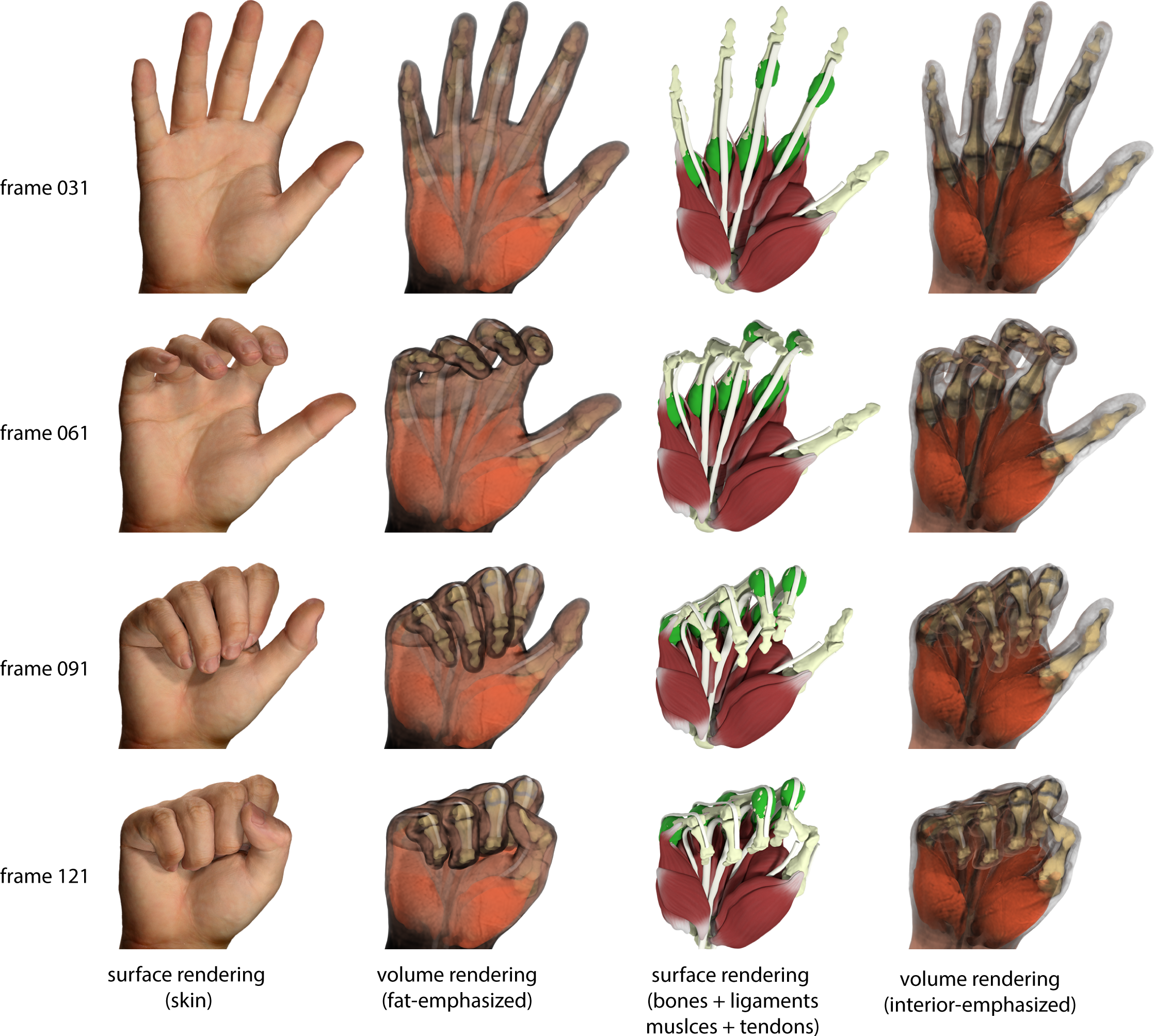}
    \caption[Comparison between standard surface rendering (left) and our volume rendering (right); ``close the fist'' motion.]
    {\textbf{Comparison between standard surface rendering (left) and our volume rendering (right); ``close the fist'' motion.}}
    \label{fig:surfaceRenderVsVolumeRender}
\end{figure}

Figure~\ref{fig:surfaceRenderVsVolumeRender} gives a comparison between surface rendering and our volume rendering. 
Surface renderings of the skin mesh and internal organs were created using Pixar RenderMan and Maya Arnold, respectively. Compared to surface rendering, our results are not photo-realistic due to the lack of textures, scattering, and global illumination. However, surface rendering cannot show the outer surface and internal structures simultaneously unless transparency is used, which diminishes the benefit of textures. 
In addition, the textures of the muscles in surface rendering do not correspond to any real data. 
Moreover, surface rendering suffers from artifacts due to penetrations between different structures, e.g., ligaments and bones, tendons and muscles. 
The penetration problems are solved by our volumetric rendering method, thanks to our priority assignment of different tissues. 
It is important to note that not only single material assignment is possible, but a linear combination of different materials 
is also easily achievable in volume rendering if necessary.

\begin{figure}
    \centering
    \includegraphics[width=0.5\textwidth]{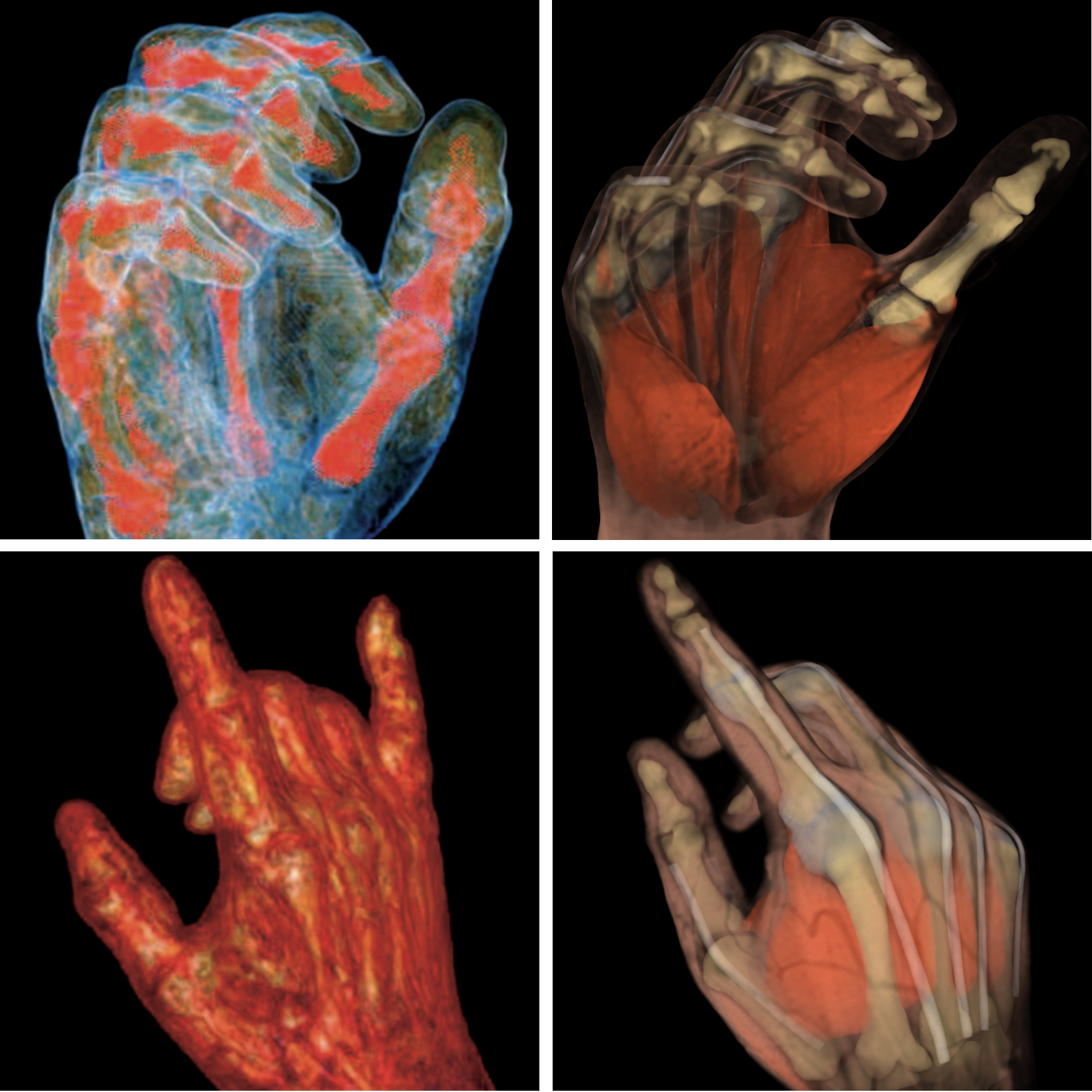}
    \caption{\textbf{Comparison of volume renders of [Rhee et al. 2010] (left) and our method (right).} (C) 2010 IEEE. Reprinted, with permission, from Rhee et al.\cite{Rhee:2010:SBVA}.}
    \label{fig:RheeVolumeRender}
\end{figure}

In Figure~\ref{fig:RheeVolumeRender}, we compared our volumetric renders to an existing quality method
for volume rendering of the human hand~\cite{Rhee:2010:SBVA}, in similar poses.
Similar to our method, they also use two rendering styles, one emphasizing internal organs, and the other emphasizing the fat tissue. 
Compared to them, in the first style, the bones, tendons and muscles
are visible in our results with clear boundaries. In the second style, in their results, 
the bones and some superficial veins (above the thumb's metacarpal) are visible, 
but the boundaries between different tissues are not as clear as with our method.

\section{Conclusion}
\label{sec:conclusion}

Starting from an MRI scan and segmented organ meshes,
we gave a volume rendering method to visualize human hand anatomy.
We utilize the known meshes to define the transfer function at each raycasting sample, 
in a manner that resolves the ambiguities in intersection areas, and removes staircasing artifacts.
We discuss the design of transfer functions for human hand anatomy, and give two families
of transfer functions, suitable for observing interior anatomy and subcutaneous fat.
We evaluated our approach in terms of rendering performance and image quality, by comparing
it to existing alternatives.

In the future, we would like to accelerate rendering using GPU computing, and potentially
achieve interactive rendering speeds.
To improve volume rendering quality, it may be beneficial to include scattering into our optical model 
and implement photo-realistic Monte-Carlo ray casting~\cite{Salama:2007:GPUMCVR,Kroes:2012:ERPRVRF,Fellner:2016:CR}.
More advanced transfer functions could be designed to utilize gradient magnitude~\cite{Levoy:1988:DSVD}. 
This could potentially visualize the boundary surfaces of veins, possibly also arteries and nerves, 
and further de-emphasize the homogeneous tissue of the fat layer.
Finally, it would be interesting to explore how to combine volumetric rendering
with other rendering techniques, such as surface rendering, non-photorealistic rendering, 
or maximum intensity projection.

\acknowledgments
{This research was sponsored in part by NSF (IIS-1911224),
USC Annenberg Fellowship to Mianlun Zheng and Bohan Wang,
Bosch Research and Adobe Research.
The authors would like to thank Shreya Bhaumik 
for her help with MRI scan segmentation, 
and Hongcheng Song for rendering help.
}

\bibliographystyle{abbrv-doi}

\bibliography{volumetricRendering}

\begin{thebibliography}{10}

\bibitem{Arens:2010:START-TF}
S.~Arens and G.~Domik.
\newblock A survey of transfer functions suitable for volume rendering.
\newblock In {\em VG@ Eurographics}, pp. 77--83, 2010.

\bibitem{Arthritis}
Arthritis.
\newblock {Hand and Wrist Anatomy website}, 2022.
\newblock
  https://www.arthritis.org/health-wellness/about-arthritis/where-it-hurts/hand-and-wrist-anatomy.

\bibitem{Bruckner:2006:ICPVR}
S.~Bruckner, S.~Grimm, A.~Kanitsar, and M.~E. Groller.
\newblock Illustrative context-preserving exploration of volume data.
\newblock {\em IEEE Transactions on Visualization and Computer Graphics},
  12(6):1559--1569, 2006.

\bibitem{Caban:2008:TBTF}
J.~J. Caban and P.~Rheingans.
\newblock Texture-based transfer functions for direct volume rendering.
\newblock {\em IEEE Transactions on Visualization and Computer Graphics},
  14(6):1364--1371, 2008.

\bibitem{Cabral:1994:VRTMH}
B.~Cabral, N.~Cam, and J.~Foran.
\newblock Accelerated volume rendering and tomographic reconstruction using
  texture mapping hardware.
\newblock In {\em Proceedings of the 1994 symposium on Volume visualization},
  pp. 91--98, 1994.

\bibitem{Cameron:1992:Shear-warp}
G.~G. Cameron and P.~E. Undrill.
\newblock Rendering volumetric medical image data on a simd-architecture
  computer.
\newblock In {\em Proceedings of the Third Eurographics Workshop on Rendering},
  pp. 135--145, 1992.

\bibitem{Cleveland}
C.~Clinic.
\newblock {Blood Vessels website}, 2022.
\newblock https://my.clevelandclinic.org/health/body/21640-blood-vessels.

\bibitem{Correa:2008:SBTF}
C.~Correa and K.-L. Ma.
\newblock Size-based transfer functions: A new volume exploration technique.
\newblock {\em IEEE transactions on visualization and computer graphics},
  14(6):1380--1387, 2008.

\bibitem{Engel:2004:RTVG}
K.~Engel, M.~Hadwiger, J.~M. Kniss, A.~E. Lefohn, C.~R. Salama, and
  D.~Weiskopf.
\newblock Real-time volume graphics.
\newblock In {\em ACM Siggraph 2004 Course Notes}, pp. 29--es. 2004.

\bibitem{Fellner:2016:CR}
F.~A. Fellner.
\newblock Introducing cinematic rendering: a novel technique for
  post-processing medical imaging data.
\newblock {\em Journal of Biomedical Science and Engineering}, 9(3):170--175,
  2016.

\bibitem{Hadwiger:2003:HQT2LVR}
M.~Hadwiger, C.~Berger, and H.~Hauser.
\newblock High-quality two-level volume rendering of segmented data sets on
  consumer graphics hardware.
\newblock In {\em IEEE Visualization, 2003. VIS 2003.}, pp. 301--308. IEEE,
  2003.

\bibitem{Kindlmann:1998:SAGTF}
G.~Kindlmann and J.~W. Durkin.
\newblock Semi-automatic generation of transfer functions for direct volume
  rendering.
\newblock In {\em IEEE Symposium on Volume Visualization (Cat. No. 989EX300)},
  pp. 79--86. IEEE, 1998.

\bibitem{Kindlmann:2003:CBTF}
G.~Kindlmann, R.~Whitaker, T.~Tasdizen, and T.~Moller.
\newblock Curvature-based transfer functions for direct volume rendering:
  Methods and applications.
\newblock In {\em IEEE Visualization, 2003. VIS 2003.}, pp. 513--520. IEEE,
  2003.

\bibitem{Kniss:2001:IVRMDT}
J.~Kniss, G.~Kindlmann, and C.~Hansen.
\newblock Interactive volume rendering using multi-dimensional transfer
  functions and direct manipulation widgets.
\newblock In {\em Proceedings Visualization, 2001. VIS'01.}, pp. 255--562.
  IEEE, 2001.

\bibitem{Kniss:2002:MDTF}
J.~Kniss, G.~Kindlmann, and C.~Hansen.
\newblock Multidimensional transfer functions for interactive volume rendering.
\newblock {\em IEEE Transactions on visualization and computer graphics},
  8(3):270--285, 2002.

\bibitem{Kroes:2012:ERPRVRF}
T.~Kroes, F.~H. Post, and C.~P. Botha.
\newblock Exposure render: An interactive photo-realistic volume rendering
  framework.
\newblock {\em PloS one}, 7(7):e38586, 2012.

\bibitem{Kruger:2003:RAY-CASTING}
J.~Kr{\"u}ger and R.~Westermann.
\newblock Acceleration techniques for gpu-based volume rendering.
\newblock In {\em Visualization Conference, IEEE}, pp. 38--38. IEEE Computer
  Society, 2003.

\bibitem{Lacroute:1994:Shear-warp}
P.~Lacroute and M.~Levoy.
\newblock Fast volume rendering using a shear-warp factorization of the viewing
  transformation.
\newblock In {\em Proceedings of the 21st annual conference on Computer
  graphics and interactive techniques}, pp. 451--458, 1994.

\bibitem{Laidlaw:1998:PVB}
D.~H. Laidlaw, K.~W. Fleischer, and A.~H. Barr.
\newblock Partial-volume bayesian classification of material mixtures in mr
  volume data using voxel histograms.
\newblock {\em IEEE transactions on medical imaging}, 17(1):74--86, 1998.

\bibitem{Levoy:1988:DSVD}
M.~Levoy.
\newblock Display of surfaces from volume data.
\newblock {\em IEEE Computer graphics and Applications}, 8(3):29--37, 1988.

\bibitem{Ljung:2016:START-TF}
P.~Ljung, J.~Kr{\"u}ger, E.~Groller, M.~Hadwiger, C.~D. Hansen, and
  A.~Ynnerman.
\newblock State of the art in transfer functions for direct volume rendering.
\newblock In {\em Computer Graphics Forum}, vol.~35, pp. 669--691. Wiley Online
  Library, 2016.

\bibitem{Lorensen:1987:MC}
W.~E. Lorensen and H.~E. Cline.
\newblock Marching cubes: A high resolution 3d surface construction algorithm.
\newblock {\em ACM siggraph computer graphics}, 21(4):163--169, 1987.

\bibitem{Lundstrom:2006:LH}
C.~Lundstrom, P.~Ljung, and A.~Ynnerman.
\newblock Local histograms for design of transfer functions in direct volume
  rendering.
\newblock {\em IEEE Transactions on visualization and computer graphics},
  12(6):1570--1579, 2006.

\bibitem{Max:1995:OMDVR}
N.~Max.
\newblock Optical models for direct volume rendering.
\newblock {\em IEEE Transactions on Visualization and Computer Graphics},
  1(2):99--108, 1995.

\bibitem{Hopkins}
J.~H. Medicine.
\newblock {Anatomy of the Hands website}, 2022.
\newblock
  https://www.hopkinsmedicine.org/health/treatment-tests-and-therapies/anatomy-of-the-hand.

\bibitem{Meissner:2000:PEVRA}
M.~Mei{\ss}ner, J.~Huang, D.~Bartz, K.~Mueller, and R.~Crawfis.
\newblock A practical evaluation of popular volume rendering algorithms.
\newblock In {\em Proceedings of the 2000 IEEE symposium on Volume
  visualization}, pp. 81--90, 2000.

\bibitem{Rezk:2008:AIT}
C.~Rezk-Salama, M.~Hadwiger, T.~Ropinski, and P.~Ljung.
\newblock Advanced illumination techniques for gpu-based volume raycasting.
\newblock {\em ACM SIGGRAPH Asia Course Notes}, 10(1508044.1508045), 2008.

\bibitem{Rhee:2010:SBVA}
T.~Rhee, J.~P. Lewis, U.~Neumann, and K.~Nayak.
\newblock Scan-based volume animation driven by locally adaptive articulated
  registrations.
\newblock {\em IEEE Transactions on Visualization and Computer Graphics},
  17(3):368--379, 2010.

\bibitem{Salama:2007:GPUMCVR}
C.~R. Salama.
\newblock Gpu-based monte-carlo volume raycasting.
\newblock In {\em 15th Pacific Conference on Computer Graphics and Applications
  (PG'07)}, pp. 411--414. IEEE, 2007.

\bibitem{Tappenbeck:2006:DBTF}
A.~Tappenbeck, B.~Preim, and V.~Dicken.
\newblock Distance-based transfer function design: Specification methods and
  applications.
\newblock In {\em SimVis}, pp. 259--274, 2006.

\bibitem{Totsuka:1993:FDVR}
T.~Totsuka and M.~Levoy.
\newblock Frequency domain volume rendering.
\newblock In {\em Proceedings of the 20th annual conference on Computer
  graphics and interactive techniques}, pp. 271--278, 1993.

\bibitem{Van:1996:DVRS3DT}
A.~Van~Gelder and K.~Kim.
\newblock Direct volume rendering with shading via three-dimensional textures.
\newblock In {\em Proceedings of 1996 Symposium on Volume Visualization}, pp.
  23--30. IEEE, 1996.

\bibitem{Wang:2019:HMA}
B.~Wang, G.~Matcuk, and J.~Barbi\v{c}.
\newblock Hand modeling and simulation using stabilized magnetic resonance
  imaging.
\newblock {\em ACM Trans. on Graphics (SIGGRAPH 2019)}, 38(4), 2019.

\bibitem{HandMRIDataset}
B.~Wang, G.~Matcuk, and J.~Barbi\v{c}.
\newblock {Hand MRI dataset}, 2020.
\newblock http://www.jernejbarbic.com/hand-mri-dataset.

\bibitem{Westover:1991:Splatting}
L.~A. Westover.
\newblock {\em Splatting: a parallel, feed-forward volume rendering algorithm}.
\newblock PhD thesis, The University of North Carolina at Chapel Hill, 1991.

\end{thebibliography}
\end{document}